\documentclass[12pt]{article}
\usepackage[a4paper]{geometry}
\usepackage{graphicx,psfrag,epsf}

\usepackage{natbib}
\usepackage{amsmath}
\usepackage{amssymb}
\usepackage{bbm}
\usepackage{booktabs}
\usepackage{multirow}
\usepackage{array}
\usepackage{rotating}
\usepackage{graphicx}
\usepackage{tcolorbox}
\usepackage{subfigure}
\usepackage{algorithm}
\usepackage{url}
\usepackage{chngcntr}

\def\bSig\mathbf{\Sigma}

\newcommand{\numclusters}{N} 
\newcommand{\numunits}{n} 

\newcommand{\link}{g} 
\newcommand{\invlink}{\link\inv} 
\newcommand{\margbetai}{\beta^{\texttt{M}}}
\newcommand{\margbeta}{\boldsymbol{\beta}^{\texttt{M}}}

\newcommand{\condbeta}{\boldsymbol{\beta}^{\texttt{C}}}
\newcommand{\margbetaest}{\widehat{\boldsymbol{\beta}}^{\texttt{M}}}
\newcommand{\condbetaest}{\widehat{\boldsymbol{\beta}}^{\texttt{C}}}
\newcommand{\margmeani}{\mu^{\texttt{M}}}

\newcommand{\margmean}{\boldsymbol{\mu}^{\texttt{M}}}

\newcommand{\margmeanlinki}{\lambda^{\texttt{M}}}
\newcommand{\margmeanlinkesti}{\widehat{\lambda}^{\texttt{M}}}

\newcommand{\margmeanlinkest}{\widehat{\boldsymbol{\lambda}}^{\texttt{M}}}
\newcommand{\margmeanlinkesttpose}{\widehat{\boldsymbol{\lambda}}^{\texttt{M}~\texttt{T}}} 
\newcommand{\condmeani}{\mu^{\texttt{C}}}

\newcommand{\condmean}{\boldsymbol{\mu}^{\texttt{C}}}

\newcommand{\logit}{\text{logit}}
\newcommand{\margbasisfun}{b^{\texttt{M}}}
\newcommand{\margbasisfunvec}{\boldsymbol{b}^{\texttt{M}}}
\newcommand{\margXbasis}{\boldsymbol{B}^{\texttt{M}}}
\newcommand{\margbasiscoef}{\boldsymbol{\alpha}^{\texttt{M}}}
\newcommand{\margbasiscoefi}{{\alpha}^{\texttt{M}}}
\newcommand{\margbasiscoefest}{\widehat{\boldsymbol{\alpha}}^{\texttt{M}}}
\newcommand{\condbasisfun}{b^{\texttt{C}}}
\newcommand{\condbasisfunvec}{\boldsymbol{b}^{\texttt{C}}}
\newcommand{\condbasiscoef}{\boldsymbol{\alpha}^{\texttt{C}}}
\newcommand{\condbasiscoefest}{\widehat{\boldsymbol{\alpha}}^{\texttt{C}}}
\newcommand{\condbasiscoefi}{{\alpha}^{\texttt{C}}}

\newcommand{\condXbasis}{\boldsymbol{B}^{\texttt{C}}}

\newcommand{\penaltyC}{\tau^{\texttt{C}}}
\newcommand{\penaltyCvec}{\mb{\penaltyC}}
\newcommand{\penaltyCest}{\widehat{\tau}^{\texttt{C}}}
\newcommand{\penaltyCestvec}{\mb{\penaltyCest}}
\newcommand{\penaltyM}{\tau^{\texttt{M}}}

\newcommand{\smoothfunM}{f^{\texttt{M}}}
\newcommand{\smoothfunC}{f^{\texttt{C}}}
\newcommand{\smoothfunMest}{\widehat{f}^{\texttt{M}}}
\newcommand{\smoothfunCest}{\widehat{f}^{\texttt{C}}}

\newcommand{\varcompC}{\boldsymbol{\theta}^{\texttt{C}}}
\newcommand{\varcompCest}{\widehat{\boldsymbol{\theta}}^{\texttt{C}}}

\newcommand{\PrecCestall}{\widehat{\boldsymbol{\mathcal{H}}}^{\texttt{C}}}

\newcommand{\PrecCestcond}{\widehat{\boldsymbol{\mathcal{H}}}_{\condbasiscoefest,\reest}^{\texttt{C}}}

\newcommand{\PrecCestcondinv}{\widehat{\boldsymbol{\mathcal{H}}}_{\condbasiscoefest,\reest}^{\texttt{C} \ -1}}

\newcommand{\PrecCestmarg}{\widehat{\boldsymbol{\mathcal{H}}}_{\penaltyCestvec,\covparamest}^{\texttt{C}}}
\newcommand{\PrecCestmarginv}{\widehat{\boldsymbol{\mathcal{H}}}_{\penaltyCestvec,\covparamest}^{\texttt{C} \ -1}}

\newcommand{\chol}{\widehat{\boldsymbol{\mathcal{L}}}}

\newcommand{\VarMf}{\boldsymbol{\mathcal{V}}_{\texttt{M}}}
\newcommand{\VarMmarg}{\boldsymbol{\mathcal{V}}^{*}_{\texttt{M}}}
\newcommand{\VarMfest}{\widehat{\VarMf}}
\newcommand{\Vfac}{\boldsymbol{V}}
\newcommand{\VarMmargest}{\widehat{\VarMmarg}}
\newcommand{\Vfacmarg}{\boldsymbol{V}^{*}}
\newcommand{\Vfacmargtpose}{\boldsymbol{V}^{*\texttt{T}}}

\newcommand{\Xmat}{\mathbf{X}}

\newcommand{\Dmat}{\mb{\mathcal{D}}}
\newcommand{\Dmatcond}{\Dmat_{\condbasiscoefest,\reest}}
\newcommand{\Dmatmarg}{\Dmat_{\penaltyCestvec,\covparamest}}

\newcommand{\rei}{u}
\newcommand{\re}{\boldsymbol{\rei}}
\newcommand{\reest}{\widehat{\re}}

\newcommand{\covmat}{\boldsymbol{\Sigma}}
\newcommand{\covmatest}{\widehat{\covmat}}
\newcommand{\covfactor}{\boldsymbol{\Lambda}}

\newcommand{\covparam}{\boldsymbol{\theta}}
\newcommand{\covparamest}{\widehat{\covparam}}

\newcommand{\redist}{F}
\newcommand{\gaussiandist}{\text{N}}
\newcommand{\redens}{\pi^{\texttt{f}}}
\newcommand{\gaussiandensity}{\phi}

\newcommand{\responsedist}{G}
\newcommand{\responsedens}{\pi^{\texttt{g}}}

\newcommand{\lik}{\mathcal{L}}
\newcommand{\likLA}{\lik_{\texttt{LA}}}
\newcommand{\mlik}{\lik^{*}}
\newcommand{\mlikLA}{\likLA^{*}}

\newcommand{\EE}{\mathbb{E}}

\newcommand{\R}{\mathbb{R}}
\newcommand{\N}{\mathbb{N}}

\newcommand{\norm}[1]{\left\lVert#1\right\rVert}

\newcommand{\ind}{\overset{ind}{\sim}}

\newcommand{\Udim}{m}

\newcommand{\paramsmalldim}{s}

\newcommand{\covxi}{x}
\newcommand{\covx}{\mb{\covxi}}
\newcommand{\covxdim}{p}
\newcommand{\covzi}{z}
\newcommand{\covz}{\mb{\covzi}}
\newcommand{\loci}{s}
\newcommand{\loc}{\mb{\loci}}
\newcommand{\region}{\mathcal{S}}

\newcommand{\tpose}{^{\texttt{T}}}
\newcommand{\inv}{^{-1}}
\newcommand{\tposeinv}{^{\texttt{T}-1}}

\newcommand{\zero}{\mb{0}}

\newcommand{\varparam}{\sigma_{\texttt{s}}}
\newcommand{\varparamind}{\sigma_{\texttt{v}}}
\newcommand{\varparamindest}{\widehat{\sigma}_{\texttt{v}}}
\newcommand{\rangeparam}{\rho}

\newcommand{\quadpointidx}{v}
\newcommand{\quadpoint}{\mb{\quadpointidx}}

\newcommand{\quadpointset}{\mathcal{Q}}

\newcommand{\weight}{\omega}
\newcommand{\quadnum}{k}

\newcommand{\gqkernel}{\phi}

\newcommand{\penaltyP}{\mathcal{P}}
\newcommand{\penaltymat}{\mb{S}}
\newcommand{\rank}{r}
\newcommand{\basisdim}{d}
\newcommand{\smoothdim}{p}

\newcommand{\datai}{y}
\newcommand{\data}{\mb{\datai}}
\newcommand{\responsei}{Y}
\newcommand{\response}{\mb{\responsei}}

\newcommand{\mb}[1]{\boldsymbol{#1}}

\def\[#1\]{\begin{equation}\begin{aligned}#1\end{aligned}\end{equation}}
\def\*[#1\]{\begin{align*}#1\end{align*}}

\usepackage{etoolbox}
\makeatletter
\patchcmd{\@makecaption}
{\parbox}
{\advance\@tempdima-\fontdimen2} 
{}{}
\makeatother

\providecommand{\keywords}[1]
{
	\small	
	\textbf{\textit{Keywords---}} #1
}

\title{Flexible Marginal Models for Dependent Data}

\usepackage{authblk}

\author{Glen McGee\thanks{glen.mcgee@uwaterloo.ca}}
\author{Alex Stringer\thanks{alex.stringer@uwaterloo.ca}}
\affil{Department of Statistics and Actuarial Science \\  University of Waterloo, Waterloo, ON, Canada}
\date{}

\begin{document}

	\maketitle
	\begin{abstract}
Models for dependent data are distinguished by their targets of inference. Marginal models are useful when interest lies in quantifying
associations averaged across a population of clusters. When the
functional form of a covariate-outcome association is unknown,  flexible
regression methods are needed to allow for potentially non-linear
relationships. We propose a novel marginal additive model (MAM) for modelling
cluster-correlated data with non-linear population-averaged associations. The
proposed MAM is a unified framework for estimation and uncertainty quantification of a marginal mean model, combined with
inference for between-cluster variability
and cluster-specific
prediction. We propose a fitting algorithm that enables efficient computation of standard errors  and corrects for estimation of penalty terms. We demonstrate the proposed methods in simulations and in application to (i) a longitudinal study of beaver foraging behaviour, and (ii) a spatial analysis of Loaloa infection in West Africa. R code for implementing the proposed methodology is available at \url{https://github.com/awstringer1/mam}.
	\end{abstract}

	%
	
	
	
	\maketitle


\section{Introduction}
\label{s:intro}

Regression models for dependent data fall into two classes defined by distinct targets of inference: 
\textit{conditional models} quantify associations within clusters, while \textit{marginal models} quantify associations averaged across a population of clusters. There exists a broad literature addressing if and when each framework is appropriate (\citealp{pendergast1996survey,neuhaus1991comparison,graubard1994regression,lindsey1998appropriateness,zeger1988models,lee2004conditional}). The choice between the two should, in principle, depend only on which association addresses the question of interest. However, most existing inference methods are specific to fitting either a conditional or marginal model, so the choice of model---and thus the choice of inferential target---is tied to the choice of inference method. As a result, analysts may end up fitting less appropriate models or estimating associations that do not correspond to the scientific question of interest.

In the case of linear covariate associations, methods for fitting conditional and marginal models are well established. Generalized linear mixed models \citep{mcculloch2008generalized}, typically fit via likelihood methods \citep{pql,approximationsnlme,maxlikglmm},  yield inference for within-cluster associations,  model within-cluster dependence, and allow prediction of cluster-specific effects. 
In contrast, the classical approach for estimating marginal associations is to solve generalized estimating equations \citep{liang1986longitudinal}, but there exist alternatives that admit full likelihood specifications. Marginalized multilevel models (MMMs; \citealp{heagerty1999marginally,heagerty2000marginalized}) permit estimation of marginal coefficients within a conditional model framework, but impose a large computational burden (see \S\ref{ss:marginalmodels}). \cite{hedeker2018note} proposed a post-hoc strategy for estimating marginal coefficients from a fitted generalized linear mixed model for binary data (\S\ref{sss:hedeker}). \citet{gory2021class} discuss when parameter estimates from a modified conditional model may have marginal interpretations.

When the functional relationship between outcomes and one or more covariates cannot be assumed linear, methods that allow for semi-parametric modelling of associations are desirable. Generalized Additive Mixed Models (GAMMs; \citealp{lin1999inference,fahrmeir2001bayesian,smoothestimation}) are likelihood-based conditional models for cluster-correlated data.
By contrast, methods for fitting marginal generalized additive models are based on estimating equations \citep{wild1996additive,berhane1998generalized,wang2005efficient,xue2010consistent,cheng2014efficient}.
While these methods do not require a correctly specified dependence structure, they cannot provide cluster-specific predictions 
or make inferences about between-cluster heterogeneity.
There are currently no likelihood-based methods for fitting non-linear marginal models.

We propose a Marginalized Additive Model (MAM) for modelling dependent data with non-linear covariate associations. The MAM admits a unified framework for estimation and uncertainty quantification of conditional and marginal mean associations and between-cluster dependence, as well as cluster-specific prediction.
We propose an estimation strategy in \S\ref{ss:estimation} that combines penalized likelihood estimation with post-hoc marginalization.
We introduce an algorithm for efficient calculation of standard errors in \S\ref{ss:standarderrors} whose computational and memory cost scales well with sample size, and which corrects for the uncertainty introduced by estimation of penalty terms and variance components. 
Simulations provide empirical evidence that the proposed procedure has generally favourable bias and coverage properties (\S\ref{s:sims}).
Finally, we demonstrate the breadth and depth of the proposed methodology in application to longitudinal study of beaver foraging behaviour in \S\ref{ss:app2} and a spatial analysis of Loaloa infection in West Africa in  \S\ref{ss:app3}.

\section{Existing Methods}
\label{s:existing}


\subsection{Notation}
\label{ss:notation}

For any $n\in\N$, let $[n] = \left\{1,\ldots,n\right\}$. Let $\datai_{ij}$ be the response, and $\responsei_{ij}$ the corresponding random variable, for the $j^{th}$ unit in the $i^{th}$ cluster, where $i\in[\numclusters]$ indexes clusters and $j\in[\numunits_i]$ 
indexes units within clusters, and let $\data = (\datai_{ij})_{i\in[\numclusters],j\in[\numunits_{i}]}$ and $\response = (\responsei_{ij})_{i\in[\numclusters],j\in[\numunits_{i}]}$. In a longitudinal analysis, for example, clusters are subjects, and units within a cluster are repeated measurements on a subject. Let $\covx_{ij}\in\R^{\covxdim}$ and $\covz_{ij}\in\R^{\Udim}$ be fixed vectors of covariates.


\subsection{Conditional Models}\label{ss:conditionalmodels}


Let $\re\sim\redist(\cdot;\covparam)$ be $\Udim$-dimensional Gaussian random effects with mean $\zero$ and covariance matrix $\covmat(\covparam)=\covfactor(\covparam)\covfactor(\covparam)\tpose$ depending on free parameters $\covparam\in\R^{\paramsmalldim}$ \citep{lme4}, such that $\redist(\cdot;\covparam) = \gaussiandist\left\{\zero;\covmat(\covparam)\right\}$. Denote by $\redens(\cdot;\covparam) = \gaussiandensity\left\{\cdot;\zero,\covmat(\covparam)\right\}$ the corresponding Gaussian density. Define the conditional mean response $\condmeani(\covx,\covz|\re)= \EE(\responsei|\covx,\covz,\re)$.
Conditional on $\re_{i}$, $\responsei_{ij}$ are assumed mutually independent with $\responsei_{ij}|\covx_{ij},\covz_{ij},\re_{i}\ind\responsedist\left\{\condmeani(\covx_{ij},\covz_{ij}|\re_{i})\right\}$ where $\responsedist (\cdot)$ is a regular response distribution \citep{smoothestimation} with densty $\responsedens(\cdot)$.
A Generalized Additive Mixed Model (GAMM) specifies
\begin{align}
\link\left\{\condmeani(\covx_{ij},\covz_{ij}|\re_{i})\right\}\ =\ \sum_{l=1}^p\smoothfunC_l(\covxi_{ijl})+ \covz_{ij}\tpose \re_{i},\label{eqn:GAMM}
\end{align}
where $\link(\cdot):\R\to\R$ is a smooth, monotone link function. The $\smoothfunC_l(\cdot)$ are unknown smooth functions represented with basis expansions,  $\smoothfunC_l(x)=\sum_{q=1}^{\basisdim_l} \condbasisfun_{lq} (x) \condbasiscoefi_{lq}$, where $\condbasisfun_{lq}(\cdot)$ are known basis functions and $\condbasiscoef = (\condbasiscoef_{l})_{l\in[p]},\condbasiscoef_{l} = (\condbasiscoefi_{q})_{q\in \basisdim_{l}}$ are $\basisdim = \sum_{l=1}^{\smoothdim}\basisdim_{l}$ coefficients to be estimated. Each $\condbasiscoef_{l}$ is associated with a penalty $\penaltyP_{\penaltyC_{l}}(\condbasiscoef_{l};\penaltymat_{l})$ where $\penaltymat_{l}$ is a $\basisdim_{l}$-dimensional fixed, known penalty matrix with rank $\rank_{l}\leq \basisdim_{l}$ and $\penaltyC_{l}>0$ is a ``smoothing'' parameter to be estimated. The full penalty $\penaltyP_{\penaltyCvec}(\condbasiscoef;\penaltymat_{1}\,\ldots,\penaltymat_{\smoothdim})=\prod_{l=1}^{\smoothdim}\penaltyP_{\penaltyC_{l}}(\condbasiscoef_{l};\penaltymat_{l})$ depends on parameters $\penaltyCvec = (\penaltyC_{l})_{i\in[\smoothdim]}$.

Contrasts of the form $\link\left\{\condmeani(\covx_{1},\covz|\re)\right\} - \link\left\{\condmeani(\covx_{2},\covz|\re)\right\}$ computed based on the 
functions $\smoothfunC_l(\cdot)$
are conditional on a particular fixed value of $\re$, which is constant only within a cluster.
The $\smoothfunC_l(\cdot)$ are therefore interpreted as cluster-specific associations.


The assumed conditional independence of $\responsei_{ij}|\covx_{ij},\covz_{ij},\re_{i}$ leads to a joint penalized likelihood,
\begin{align}
\lik(\condbasiscoef,\re,\penaltyCvec,\covparam;\data) = \left\{\prod_{i=1}^{\numclusters}\prod_{j=1}^{n_j}\lik_{ij}(\condbasiscoef,\re_{i},\covparam;\datai_{ij})\right\}\times\penaltyP_{\penaltyCvec}(\condbasiscoef;\penaltymat_{1}\,\ldots,\penaltymat_{\smoothdim}), \label{eqn:marglikglmm}
\end{align}
where $\lik_{ij}(\condbasiscoef,\re_{i},\covparam;\datai_{ij}) = \responsedens\left\{\condmeani(\covx_{ij},\covz_{ij}|\re_{i})\right\}\redens(\re_{i};\covparam)$, and hence a penalized marginal likelihood, $$\mlik(\penaltyCvec,\covparam;\data) = \int\lik(\condbasiscoef,\re,\penaltyCvec,\covparam;\data)d\redist(\re;\covparam)d\condbasiscoef.$$Denote by $\mlikLA(\penaltyCvec,\covparam;\data)$ a Laplace approximation to $\mlik(\penaltyCvec,\covparam;\data)$. Inference for the parameters $(\condbasiscoef,\penaltyCvec,\covparam)$ follows from first-order approximations to the sampling distributions of $(\penaltyCestvec,\covparamest) = \text{argmax} \ \mlikLA(\penaltyCvec,\covparam;\data)$ and $(\condbasiscoefest,\reest) = \text{argmax} \ \lik(\condbasiscoef,\re,\penaltyCestvec,\covparamest;\data)$ \citep{wood2011fast,smoothestimation}.

\subsection{Marginal Models}\label{ss:marginalmodels}


Define the \emph{marginal} (or population-averaged) mean, $\margmeani(\covx)\equiv \EE(\responsei|\covx)$. A marginal model is one in which $\margmeani(\covx)$ is modelled directly, and this is desirable when interest lies in $\margmeani(\covx)$ and contrasts involving it.
However, the $\responsei_{ij}$ are not marginally independent. This makes it challenging to define their joint distribution, and hence directly specify a full likelihood for
any parameters upon which $\margmeani(\covx)$ depends.

The marginal mean can always be recovered from a conditional model:
\begin{align}
\margmeani(\covx)\ =\ \int \invlink\left\{\condmeani(\covx,\covz|\re)\right\}d\redist(\re; \covparam).  \label{eqn:inteq}
\end{align} 
This relationship (\ref{eqn:inteq}) can be exploited to make likelihood-based inferences for $\margmeani(\covx)$ in cases when likelihood inferences are available for the conditional model.

A Marginalized Multilevel Model (MMM; \citealt{heagerty1999marginally,heagerty2000marginalized}) assumes a \emph{linear} marginal model $\link\left\{\margmeani(\covx)\right\} =\covx\tpose\margbeta$, and pairs it with a conditional dependence model,
\begin{align}
\link\left\{\condmeani(\covx_{ij},\covz_{ij}|\re_{i})\right\} \ =\ \Delta(\covx_{ij})\ +\ \covz_{ij}\tpose \re_i,  \label{eqn:MMMcond} %
\end{align}
where $\Delta(\covx_{ij})$ are implicitly defined by (\ref{eqn:inteq}).
The MMM therefore specifies a likelihood for $(\margbeta,\covparam)$ through (\ref{eqn:MMMcond}), (\ref{eqn:marglikglmm}), and (\ref{eqn:inteq}).
However, in order to calculate the likelihood contribution $\lik(\margbeta,\re_{i};\datai_{ij})$ for each $\datai_{ij}$, for any given $\margbeta$, $\Delta(\covx_{ij})$ must first be computed by solving (\ref{eqn:inteq}). 
That $n=\sum_{i=1}^{\numclusters}\numunits_{i}$ integral approximations and non-linear equation solutions are required \emph{per evaluation} of the marginal likelihood (\ref{eqn:marglikglmm}) in an MMM leads to computational burden,
limiting the practical application of MMMs.

\subsection{Post-Hoc Marginalization} 
\label{sss:hedeker}
Focusing on linear models,
\cite{hedeker2018note} proposed a strategy for estimating marginal coefficients $\margbeta$ after fitting a conditional model.
\citet{hedeker2018note}  jointly adopts the following pair of models:
 \begin{align}
 \link\left\{\margmeani(\covx_{ij})\right\}\ &=\ \covx_{ij}\tpose\margbeta, \label{eqn:HedekerMarg}\\
 \link\left\{\condmeani(\covx_{ij},\covz_{ij}|\re_{i})\right\}&=\ \covx_{ij}\tpose \condbeta+ \covz_{ij}\tpose \re_i.  \label{eqn:GLMM}
 \end{align}
Since by (\ref{eqn:inteq}), (\ref{eqn:HedekerMarg}) and (\ref{eqn:GLMM}) cannot hold simultaneously, we view the combination of (\ref{eqn:HedekerMarg}) and (\ref{eqn:GLMM}) as replacing $\Delta(\covx_{ij})$ in (\ref{eqn:MMMcond}) with a linear approximation $\Delta(\covx_{ij})\approx\covx_{ij}\tpose \condbeta$. Define the $n\times p$ design matrix of covariates $\Xmat=[\Xmat_{1\text{$\cdot$}}\tpose,\dots,\Xmat_{N\text{$\cdot$}}\tpose]\tpose$, the $n_i\times p$ covariate matrix for the $i^{th}$ cluster $\Xmat_{i\text{$\cdot$}}=[\covx_{i1},\dots,\covx_{in_i}]\tpose$, and the linked marginal mean $\margmeanlinki_{ij}\equiv \link\left\lbrace \margmeani(\covx_{ij})\right\rbrace$. Estimation and inference for (\ref{eqn:HedekerMarg}) proceeds by: i) fitting (\ref{eqn:GLMM}),
yielding estimates $\condbetaest$ and $\covparamest$; ii) setting $\Delta(\covx_{ij}) = \covx_{ij}\tpose\condbetaest$ and $\covmat=\covmat(\covparamest)$ in (\ref{eqn:inteq}) and approximating the integral using Gauss-Hermite quadrature once for each $i\in[\numclusters],j\in[\numunits_{i}]$, obtaining approximate estimates $\margmeanlinkesti(\covx_{ij})$; and iii) substituting these estimates into (\ref{eqn:HedekerMarg}) and solving, yielding $\margbetaest = (\Xmat\tpose\Xmat)\inv\Xmat\tpose\margmeanlinkest$ where $\margmeanlinkest=(\margmeanlinkesttpose_{1\text{$\cdot$}},\dots,\margmeanlinkesttpose_{N\text{$\cdot$}})\tpose$ and $\margmeanlinkest_{i\text{$\cdot$}}=(\margmeanlinkesti(\covx_{i1}),\dots, \margmeanlinkesti(\covx_{in_i}))\in\R^{n}$.

The post-hoc marginalization of a conditional model is a compelling strategy for fitting a marginal model, having the full benefits of a unified likelihood-based framework for fitting both models simultaneously, while  circumventing the computational burden of MMMs.

\section{Flexible Marginal Models for Dependent Data}


\label{s:methods}

\subsection{Marginal Additive Models}\label{ss:mam}
The Marginal Additive Model (MAM) is:
\begin{align}
  \link\left\{\margmeani(\covx_{ij})\right\}\ &=\ \smoothfunM(\covx_{ij}), \label{eqn:flexMarg} \\
  \link\left\{\condmeani(\covx_{ij},\covz_{ij}|\re_{i})\right\}&=\ \Delta(\covx_{ij}) + \covz_{ij}\tpose \re_{i},  \label{eqn:flexCond}
\end{align}
where $\smoothfunM(\covx_{ij})\equiv\sum_{l=1}^p \smoothfunM_l(\covxi_{ijl})$ is an additive function. Like in a MMM (\ref{eqn:MMMcond}), $\Delta(\covx_{ij})$ is implicitly defined by (\ref{eqn:flexMarg}) and (\ref{eqn:flexCond}) according to (\ref{eqn:inteq}).



\subsection{Estimation}
\label{ss:estimation}


Inference for $\smoothfunM(\covx_{ij})$ in a MAM is facilitated by a further additive approximation $\Delta(\covx)\approx\smoothfunC(\covx)=\sum_{l=1}^p\smoothfunC_l(\covxi_{l})$ as defined in \S\ref{ss:conditionalmodels}. The full method for fitting the MAM is described in Algorithms 1 -- 3 
in the supplementary material. We provide an overview here.

We estimate $\smoothfunC(\covx)$ by fitting a GAMM (\ref{eqn:GAMM}) using penalized likelihood as described in \S\ref{ss:conditionalmodels}. Write $\smoothfunC(\Xmat)= {\condXbasis} \condbasiscoef$ where $\condXbasis\in\R^{n\times \sum_l^p Q_l}$ has rows $\condXbasis_{ij}=[{{\condbasisfunvec}_{ij1}}\tpose,\dots,{\condbasisfunvec_{ijp}}\tpose]$ where $\condbasisfunvec_{ijl}=[\condbasisfun_{l1} (\covxi_{ijl}),\dots,\condbasisfun_{lQ_l} (\covxi_{ijl}) ]\tpose$. 
Denote the corresponding estimates by $\smoothfunCest(\covx_{ij}) = {\condXbasis_{ij}} \condbasiscoefest$.

Next, for each $i\in[\numclusters],j\in[\numunits_{i}]$, with $\margmeanlinki_{ij}\equiv \link\left\lbrace \margmeani(\covx_{ij})\right\rbrace$ as in \S\ref{sss:hedeker}, approximate:
\begin{equation}\label{eqn:ghqestimate}
\margmeanlinkesti(\covx_{ij}) = \link\left[ \sum_{\quadpoint\in\quadpointset(\Udim,\quadnum)}\invlink\left\{\smoothfunCest(\covx_{ij}) + \covz_{ij}\tpose\covfactor(\covparamest)\quadpoint\right\}\gqkernel(\quadpoint)\weight_{\quadnum}(\quadpoint)\right].
\end{equation}
Here $\gqkernel:\R^{\Udim}\to\R$ is an $\Udim$-dimensional standard Gaussian density, and $\quadpointset(\Udim,\quadnum)\subset\R^{\Udim}$ and $\weight_{\quadnum}:\quadpointset(\Udim,\quadnum)\to\R^{+}$ are the nodes and weights, respectively, from an $\Udim$-dimensional product Gauss-Hermite quadrature rule.

Finally, we represent $\smoothfunM_l(x)$ via further basis expansions: $\smoothfunM_l(x)=\sum_{q=1}^{Q_l} \margbasisfun_{lq} (x) \margbasiscoefi_{lq}$, where $\margbasisfun_{lq}(\cdot)$ are known basis functions and $\margbasiscoef=(\margbasiscoefi_{lq})_{l\in[p],q\in Q_{l}}$ are coefficients. Write $\smoothfunM(\Xmat)= {\margXbasis} \margbasiscoef$ where $\margXbasis\in\R^{n\times \sum_l^p Q_l}$ has rows $\margXbasis_{ij}=[{{\margbasisfunvec}_{ij1}}\tpose,\dots,{\margbasisfunvec_{ijp}}\tpose]$ where $\margbasisfunvec_{ijl}=[\margbasisfun_{l1} (\covxi_{ijl}),\dots,\margbasisfun_{lQ_l} (\covxi_{ijl}) ]\tpose$. Let $\margmeanlinkest = \left( \margmeanlinkesti(\covx_{ij})\right)_{i\in[\numclusters],j\in[\numunits_{i}]}$, and estimate:
\begin{align}\label{eqn:mamls}
\margbasiscoefest&= \text{argmin}\norm{\margmeanlinkest - \margXbasis\margbasiscoef} = \left({\margXbasis}\tpose  \margXbasis \right)^{-1}{\margXbasis}\tpose  \margmeanlinkest.
\end{align}


In the linear case, \cite{hedeker2018note} first fits a conditional model via maximum likelihood, and in the final step, solves equation (\ref{eqn:HedekerMarg}) for $\margbeta$ and plugs in estimates as necessary, yielding a least-squares-style estimator similar to (\ref{eqn:mamls}). In the additive setting, we fit a conditional model via penalized likelihood, and it is tempting, by analogy, to minimize a penalized least squares criterion to estimate $\margbasiscoefest$. However, this is not necessary: the $\margmeanlinkesti(\covx_{ij})$ are deterministic functions of the (stochastic) estimated means and variance components, rather than observed responses. Smoothness has already been imposed by estimating $\margmeani(\covx_{ij})$ via penalized likelihood. Thus we simply project $\margmeanlinkesti(\covx_{ij})$ onto the column space of the basis functions via ordinary least squares. Figure B.3 in the Supplementary Material gives empirical evidence in support of the claim that a second round of penalization is unnecessary. 


The final marginal function estimates are $\smoothfunMest(\Xmat)= {\margXbasis} \margbasiscoefest$. The structure of $\margmeanlinkest$ is further exploited to obtain estimates of pointwise standard errors of $\smoothfunMest(\Xmat)$.






\subsection{Standard Errors}\label{ss:standarderrors}

The MAM estimates $\smoothfunMest(\Xmat)$ are continuous functions of $(\condbasiscoefest,\reest,\penaltyCestvec,\covparamest)$, and we compute approximate standard errors via the Delta method. \citet{hedeker2018note} adopts a similar approach, requiring that a joint \emph{covariance} matrix for $(\condbeta,\covparam)$ is available. This is a reasonable requirement in the linear setting they consider. 

In the additive setting, more effort is required to obtain standard errors in a computationally efficient manner. Let $\PrecCestcond=-\partial^{2}_{\condbasiscoef,\re}\log\lik(\condbasiscoefest,\reest,\penaltyCestvec,\covparamest;\data)$, with dimension $(\basisdim + \numclusters\Udim)\times(\basisdim + \numclusters\Udim)$. Holding $(\penaltyCvec,\covparam) = (\penaltyCestvec,\covparamest)$ fixed, a na\"ive expression for an approximate $n\times n$ variance matrix $\VarMfest \approx \text{Var}\left(\smoothfunMest(\Xmat)|\penaltyCestvec,\covparamest\right)$, is
\begin{equation}\label{eqn:naivevariance}
\VarMfest = \margXbasis\left({\margXbasis}\tpose  \margXbasis \right)\inv{\margXbasis}\tpose\left( \Dmatcond\PrecCestcondinv\Dmatcond\tpose\right)\margXbasis\left({\margXbasis}\tpose  \margXbasis \right)\inv{\margXbasis}\tpose,
\end{equation}
where $\Dmatcond = \partial_{\condbasiscoef,\re}\margmeanlinkest$ is of dimension $n\times(\basisdim + \numclusters\Udim)$. 

Attempting to directly use (\ref{eqn:naivevariance}) for inference presents two computational challenges:
\begin{enumerate}
  \item[(a)] $\text{dim}\left(\VarMfest\right) = n\times n$, so computing and storing $\VarMfest$ is computationally challenging,
  \item[(b)] Fixing $(\penaltyCvec,\covparam) = (\penaltyCestvec,\covparamest)$ can lead to undercoverage of the approximate intervals.
\end{enumerate}



Neither challenge is present when fitting the linear models considered by \citet{hedeker2018note}. Challenge (a) is avoided because in that setting, interest lies in the regression coefficients, not the entire fitted curve, a standard difference between linear and nonlinear regression modelling. Hence there are only $p\ll n$ (say) standard errors required, and only $p\times p$ matrices to compute and store, where $p$ does not increase with $n$. 

Challenge (b) is absent because in the linear case, the regression coefficients and variance components are \emph{jointly} estimated by maximizing a marginal likelihood in which the random effects have been averaged over, and approximate variances computed using the Hessian of this marginal likelihood properly account for uncertainty in both the regression coefficients and variance components. In the additive setting, the variance components $\covparamest$ are estimated by marginal likelihood, but the basis function coefficients $\condbasiscoefest$ are estimated by penalized (joint) likelihood. There is no guarantee that the joint Hessian $\PrecCestall = -\partial^{2}_{\condbasiscoefest,\reest,\penaltyCestvec,\covparamest}\log\lik(\condbasiscoefest,\reest,\penaltyCestvec,\covparamest;\data)$ will be positive definite at the specific point $(\condbasiscoefest,\reest,\penaltyCestvec,\covparamest)$, a point that is not the mode of $\log\lik(\condbasiscoef,\re,\penaltyCvec,\covparam;\data)$. The strategy of using this joint mode and Hessian for inference is well documented to be unfavourable compared to strategies that use both penalized and marginal likelihoods in the proposed manner \citep{kassandsteffy,smoothestimation,blockupdate,gmrfmodels}.

We address challenge (a) by introducing an efficient algorithm for computing the $n$ diagonal values $\VarMfest_{ii},i\in[n]$ upon which pointwise confidence intervals for $\smoothfunM(\covx_{i})$ are based, without ever computing or storing any $n\times n$ matrices. We address challenge (b) by correcting these computed approximate marginal variances for the uncertainty introduced by estimating $(\penaltyCvec,\covparam)$ using a variant of the methods of \citet{kassandsteffy,smoothestimation}.

The independence of $\re=(\rei_i)_{i\in[\numclusters]}$ means that the $\numclusters\times\numclusters$ block of $\PrecCestcond$ containing pairwise second derivatives $-\partial^{2}_{\rei_{i}\rei_{j}}\log\lik(\condbasiscoefest,\reest,\penaltyCestvec,\covmatest;\data), i,j\in[\numclusters]$ is diagonal. This means that $\PrecCestcond$ has only $O(\numclusters)$ non-zero entries despite having $O(\numclusters^{2})$ total entries, which we exploit to reduce memory cost. We compute the $(\basisdim + \numclusters\Udim)\times(\basisdim + \numclusters\Udim)$ Cholesky decomposition $\chol$ of $\PrecCestcond=\chol\chol\tpose$ using fill-reducing permutations \citep[ \S2.4]{gmrf}, leading to a memory-efficient sparse representation with only $O(\numclusters)$ non-zero entries to be stored, and efficient solution of linear systems. We write $\VarMfest \equiv \Vfac\Vfac\tpose$, where
$$
\Vfac\tpose = \left\{ \chol\tposeinv\Dmatcond\tpose\right\}\margXbasis\left({\margXbasis}\tpose  \margXbasis \right)\inv{\margXbasis}\tpose,
$$
and $\Vfac$ has dimension $n\times(\basisdim + \numclusters\Udim)$. We compute the product $\chol\tposeinv\Dmat\tpose$ efficiently using a sparse triangular solve. We then compute required pointwise variances as $$(\VarMfest)_{ii} = \sum_{j=1}^{\basisdim + \numclusters\Udim}(\Vfac_{ij})^{2}, i\in[n].$$ Crucially, this avoids storing dense $n\times n$ matrices, thus addressing challenge (a). Further, the $n$ required column sums can be computed in parallel for large $n$.


Having treated $\penaltyCestvec$ and $\covparamest$ as fixed, we address challenge (b) via a variant of the approach of \citet{kassandsteffy,smoothestimation}. Let $\PrecCestmarg = -\partial^{2}_{\penaltyCvec,\covparam}\log\mlikLA(\penaltyCvec,\covparam;\data)$, of dimension $(\smoothdim + \paramsmalldim)\times(\smoothdim + \paramsmalldim)$, $\Dmatmarg = \partial_{\penaltyCvec,\covparam}\margmeanlinkest$, of dimension $n\times(\smoothdim + \paramsmalldim)$, and $$\VarMmargest = \margXbasis\left({\margXbasis}\tpose  \margXbasis \right)\inv{\margXbasis}\tpose\left( \Dmatmarg\PrecCestmarginv\Dmatmarg\tpose\right)\margXbasis\left({\margXbasis}\tpose  \margXbasis \right)\inv{\margXbasis}\tpose \equiv \Vfacmarg\Vfacmargtpose.$$ We compute $(\VarMmargest)_{ii}$ in the same way as $(\VarMfest)_{ii}$. The final pointwise variance approximation is $\text{Var}(\smoothfunM(\covx_{i})) \approx (\VarMfest)_{ii} + (\VarMmargest)_{ii}$. Approximate $(1-\alpha)$ confidence intervals are obtained as $$\smoothfunMest(\covx_{i})\pm z_{\alpha/2}\left\{(\VarMfest)_{ii} + (\VarMmargest)_{ii}\right\}^{1/2}.$$ The empirical coverage properties of these intervals is assessed in \S\ref{s:sims}. Full details are prescribed in Algorithm 3 in the Supplementary Materials.



\subsection{Implementation}\label{ss:implementation}

Of the quantities described in \S\ref{ss:standarderrors}, several are nontrivial to compute and require further comment. We compute $\chol\tposeinv\Dmat\tpose$ using the \texttt{Cholesky} and \texttt{solve} functions in the \texttt{Matrix} package \citep{matrix} in the \texttt{R} language \citep{rlanguage}, which implements algorithms described by \citet{matrixalgorithm1} and \cite{matrixalgorithm2} for LDL decompositions with fill-reducing permutations and sparse system solves as used in \S\ref{ss:standarderrors}.

We compute the Jacobian $\Dmat=\partial_{\condbasiscoef,\re,\penaltyCvec,\covmat}\margmeanlinkest$ using automatic differentiation, implemented in the \texttt{TMB} package \citep{kristensen2015tmb}, and to obtain $\Dmatcond$ and $\Dmatmarg$ by indexing the appropriate columns of $\Dmat$. The explicit use of sparse matrix and vector algebra exploits the fact that many entries of $\Dmat$ are zero by construction.

\section{Simulations}
\label{s:sims}

\subsection{Setup}
\label{ss:simsetup}
We investigate the bias and coverage properties of the proposed method empirically via a simulation study. We generated $R=500$ datasets as follows. We set the number of clusters $\numclusters \in \{100,200\}$ and the number of units within clusters $\numunits_i\in \{10,20\}$ $\forall i \in \{1,\dots,\numclusters\}$. For each unit $j \in  \{1,\dots,\numunits_i\}$, we generated three covariates, $\covx_{ij}=(x_{ij1},x_{ij2},x_{ij3})\tpose$ where $x_{ijp}\sim \text{Unif}(-1,1)$ for $p=1, 2, 3.$ We generated binary outcomes $y_{ij}$ according to the following marginal mean model:
\begin{align}
	\text{logit}\left\{\margmeani(\covx_{ij})\right\}\ &=\ \smoothfunM_1(x_{ij1}) +\smoothfunM_2(x_{ij2}) +\beta_3 \cdot x_{ij3}, \label{eqn:genMarg} 
\end{align}
where $\smoothfunM_1(\cdot)$ and $\smoothfunM_2(\cdot)$ are smooth functions (see supplementary material for specific functions),  and a random intercepts and slopes dependence structure:
\begin{align*}
	\text{logit}\left\{\condmeani(\covx_{ij}|\rei_{i})\right\}&=\ \Delta(\covx_{ij}) + \rei_{i0}+\rei_{i1} \cdot x_{ij3}, \\  \left[\begin{array}{c}
		\rei_{i0} \\ \rei_{i1}\\
	\end{array}\right]&\sim MVN\left(\left[\begin{array}{c}
0 \\ 0\\
	\end{array}\right], 
\left[\begin{array}{cc}
	\sigma_0^2 & \rho \sigma_0 \sigma_1 \\
	\rho \sigma_0 \sigma_1 & \sigma_1^2 \\
\end{array}\right]\right) 
\end{align*}
where we set $\sigma_0=2$, $\sigma_1=1$, $\rho=0.5$, and $\beta_3=0$. (See Supplementary Material for extended simulations under random intercepts dependence structure.)

To each of the $R$ datasets, we fit a GAM (via the \texttt{gam} function in the \texttt{mgcv} package \citealp{wood2015package}), a GAMM (via the \texttt{gamm4} function of the  \texttt{gamm4} package \citealp{wood2017package}), and the proposed MAM (software included as supplementary material). In each model, the smooth functions  $\smoothfunM_1(\cdot)$ and $\smoothfunM_2(\cdot)$ were treated as unknown and were estimated from the data via thin plate regression splines \citet{thinplatesplines}. The GAM estimates a marginal mean,
but does not account for the cluster-correlated structure. The GAMM explicitly models the random effects, but does not estimate the marginal mean. The MAM both estimates a marginal mean and explicitly models the random effects. 

We report mean bias and 95\% confidence interval coverage for estimates of $\smoothfunM_1(\cdot)$ and $\smoothfunM_2(\cdot)$ on a grid of evenly spaced covariate values; bias for estimates of variance components $\varcompC=(\sigma_0,\sigma_1,\rho)^T$, and root mean squared error of prediction (RMSEP) for the random intercepts $\rei_{i0}$ and  slopes $\rei_{i1}$.

\subsection{Results}
\label{ss:results}

 Figure B.1 in the supplementary material shows distributions of estimates and 95\% confidence interval coverages  for  $f_1(\cdot)$ and $f_2(\cdot)$ on a grid of evenly spaced covariate values, and Table \ref{tab:sim2}  shows bias and interval coverage averaged across the grid of covariates. The GAM and MAM both yielded similar estimates of $f_1(\cdot)$ and $f_2(\cdot)$, and showed little bias on average across the grid of covariates. The GAM had poor coverage, however, as it erroneously treated observations as independent, while the MAM achieved nominal coverage.

\begin{table}
\caption{Simulation results. Bias and 95\% confidence interval coverage (Cvg; in \%) for estimates of $f^{\texttt{M}}_1(x_1)$ and $f^{\texttt{M}}_2(x_2)$ averaged over a grid of evenly spaced values of $x_1$ and $x_2$, bias in estimates of variance components $\varcompC=(\sigma_0,\sigma_1,\rho)^T=(2,1,0.5)^T$, root mean squared error of prediction (RMSEP) for random intercepts $\rei_{i0}$ and slopes $\rei_{i1}$.  \label{tab:sim2}  }
	\centering
	\begin{tabular}{lll r@{\extracolsep{5pt}}r r@{\extracolsep{5pt}}r r@{\extracolsep{5pt}}r@{\extracolsep{5pt}}r r@{\extracolsep{5pt}}r}
		\toprule
		&&&  \multicolumn{2}{c}{$\widehat{f^{\texttt{M}}_1}(\cdot)$ } & \multicolumn{2}{c}{$\widehat{f^{\texttt{M}}_2}(\cdot)$ }  & \multicolumn{3}{c}{ Bias(${\varcompCest}$)}& \multicolumn{2}{c}{ RMSEP}  \\
		\cmidrule(l{2pt}r{2pt}){4-5}   \cmidrule(l{2pt}r{2pt}){6-7} \cmidrule(l{2pt}r{2pt}){8-10} \cmidrule(l{2pt}){11-12} 
		Model  & $\numclusters$& $\numunits_i$  & Bias & Cvg & Bias & Cvg &  $\sigma_0$ & $\sigma_1$ & $\rho$ & $\rei_{i0}$& $\rei_{i1}$\\
		\midrule
		GAM & 100 & 10 & -0.06 & 92 & -0.03 & 94 & -- & -- & -- & -- & -- \\
			&  	  & 20 & -0.03 & 90 & -0.02 & 91 & -- & -- & -- & -- & -- \\
			& 200 & 10 & -0.04 & 93 & -0.02 & 93 & -- & -- & -- & -- & -- \\
			&     & 20 & -0.03 & 90 & -0.04 & 88 & -- & -- & -- & -- & -- \\
		\\[-1.8ex]   
		GAMM & 100 & 10 & -- & -- & -- & -- & -0.11 & -0.28 & -0.03 & 3.40 & 1.27 \\
			 &     & 20 & -- & -- & -- & -- & -0.04 & -0.12 & -0.08 & 3.63 & 1.44 \\
			 & 200 & 10 & -- & -- & -- & -- & -0.10 & -0.27 &  0.00 & 3.43 & 1.27 \\
			 &     & 20 & -- & -- & -- & -- & -0.06 & -0.14 & -0.09 & 3.61 & 1.41 \\
		\\[-1.8ex]
		MAM & 100 & 10 & -0.02 & 97 & 0.07 & 96 & -0.08 & -0.26 & -0.04 & 3.40 & 1.28 \\
			&     & 20 &  0.00 & 96 & 0.08 & 96 & -0.01 & -0.10 & -0.09 & 3.63 & 1.44 \\
			& 200 & 10 &  0.00 & 96 & 0.08 & 94 & -0.08 & -0.27 &  0.00 & 3.42 & 1.27 \\
			&     & 20 &  0.00 & 97 & 0.04 & 95 & -0.05 & -0.13 & -0.09 & 3.61 & 1.41 \\
		\hline
	\end{tabular}
\end{table}

Table \ref{tab:sim2} also reports bias in estimates of the variance components $\varcompC$ as well as RMSEP for the random effects. The GAMM and the MAM performed similarly,  estimating $\sigma_0$ with little bias, but exhibiting some downward bias in estimating $\sigma_1$ when cluster size was small.  Both yielded the same prediction accuracy.

The MAM is the only method able to both (a) conduct valid inference on the marginal exposure-response curves $\smoothfunM_1(\cdot)$ and $\smoothfunM_2(\cdot)$ and (b) estimate variance components and predict random effects.


\section{Applications}
\label{s:app}	
We demonstrate the breadth and depth of the proposed MAM on a longitudinal study from ecology  and a spatial study from epidemiology.

\subsection{Seasonality in Beaver Foraging Behaviour}
\label{ss:app2}
\cite{lodberg2021size} investigated seasonal trends in foraging behaviour in a sample of  beavers measured over time \citep{lodbergholm2021}. The sample includes a total of $\numunits=\sum_{i=1}^\numclusters \numunits_{i}=3,311$ observations of $\numclusters=34$ beavers while foraging, and the outcome of interest is an indicator of whether they foraged on trees/shrubs (vs. aquatic vegetation/herbs/grasses; 1/0). Interest lies in sex differences in seasonal trends in foraging behaviour. The original analysis fit a GAMM with a logit link to flexibly model seasonal trends in the behaviour of the $34$ particular beavers studied, while accounting for repeated measurements on those beavers. 
In contrast, when the aim of the analysis is to infer the association between environmental changes and the average foraging behaviour of the population of beavers, the following MAM is appropriate: 
\begin{align*}\label{eqn:bvmodel}
			\text{logit}\left\{\margmeani(\covx_{ij},\texttt{time}_{ij})\right\}\ &=\ \covx_{ij}\tpose\margbeta + \texttt{I}(\texttt{male}_{ij}) \smoothfunM_{\texttt{m}}\{\texttt{time}_{ij}\} + \left[1- \texttt{I}(\texttt{male}_{ij})\right] \smoothfunM_{\texttt{f}}\{\texttt{time}_{ij}\} , \\
		\text{logit}\left\{\condmeani(\covx_{ij},\texttt{time}_{ij}|\re_{i})\right\}&=\  \Delta(\covx_{ij},\texttt{time}_{ij})+ \rei_i, \\ 	\rei_i&\ind\text{Normal}\left\{0,\texttt{I}(\texttt{male}_{ij})\sigma_{m}^{2}+ \left[1- \texttt{I}(\texttt{male}_{ij})\right]\sigma_{f}^{2}\right\}, 
	\end{align*}
where $\texttt{I}(\texttt{male}_{ij})$ is an indicator of male sex, $\texttt{time}_{ij}$ is day of the year, and $\covx_{ij}$ includes indicators for year (categorical: 2000, 2006, 2007).

\begin{figure}[htbp!]
	\centering

	\subfigure[Female]{\includegraphics[width=0.49\linewidth]{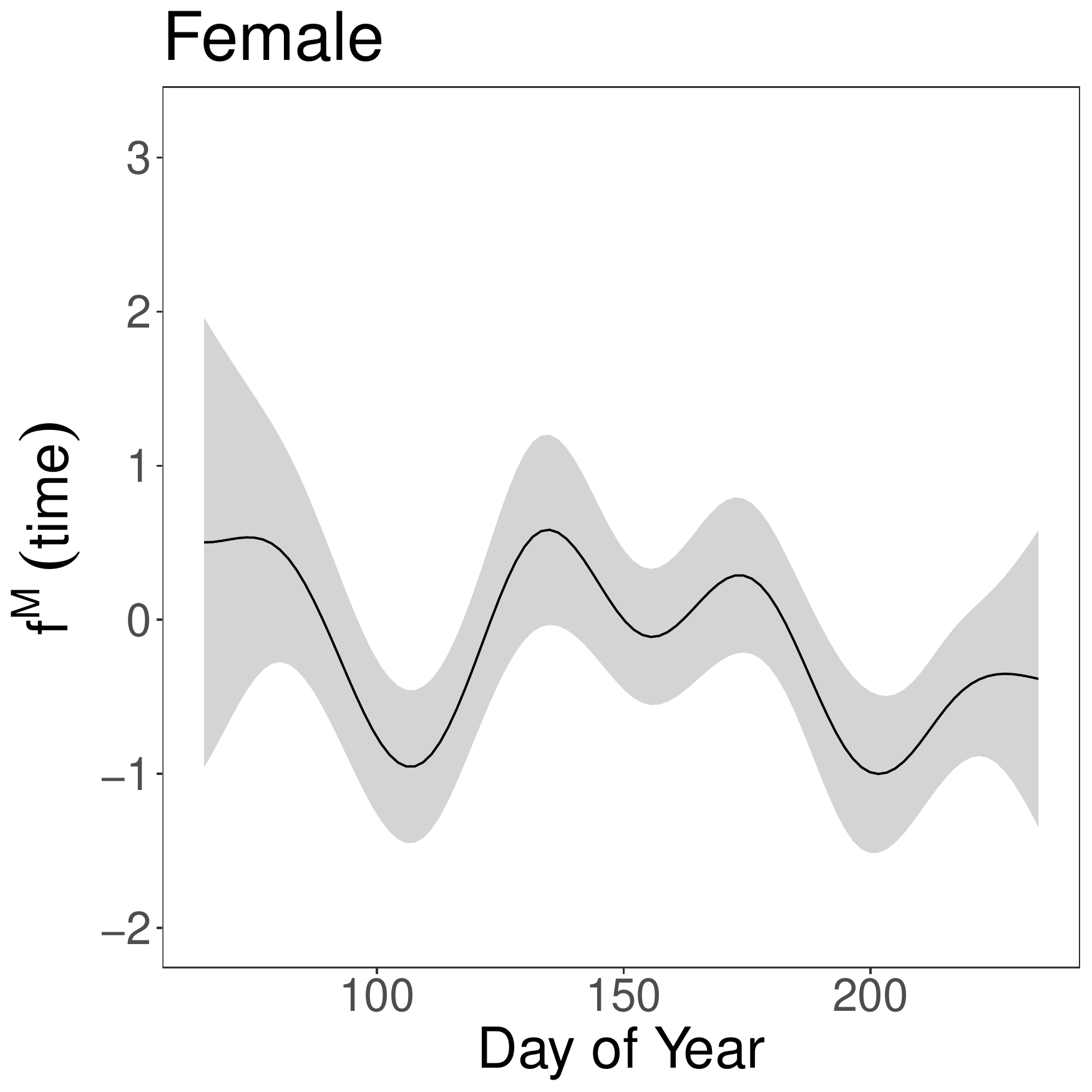}}
	\subfigure[Male]{\includegraphics[width=0.49\linewidth]{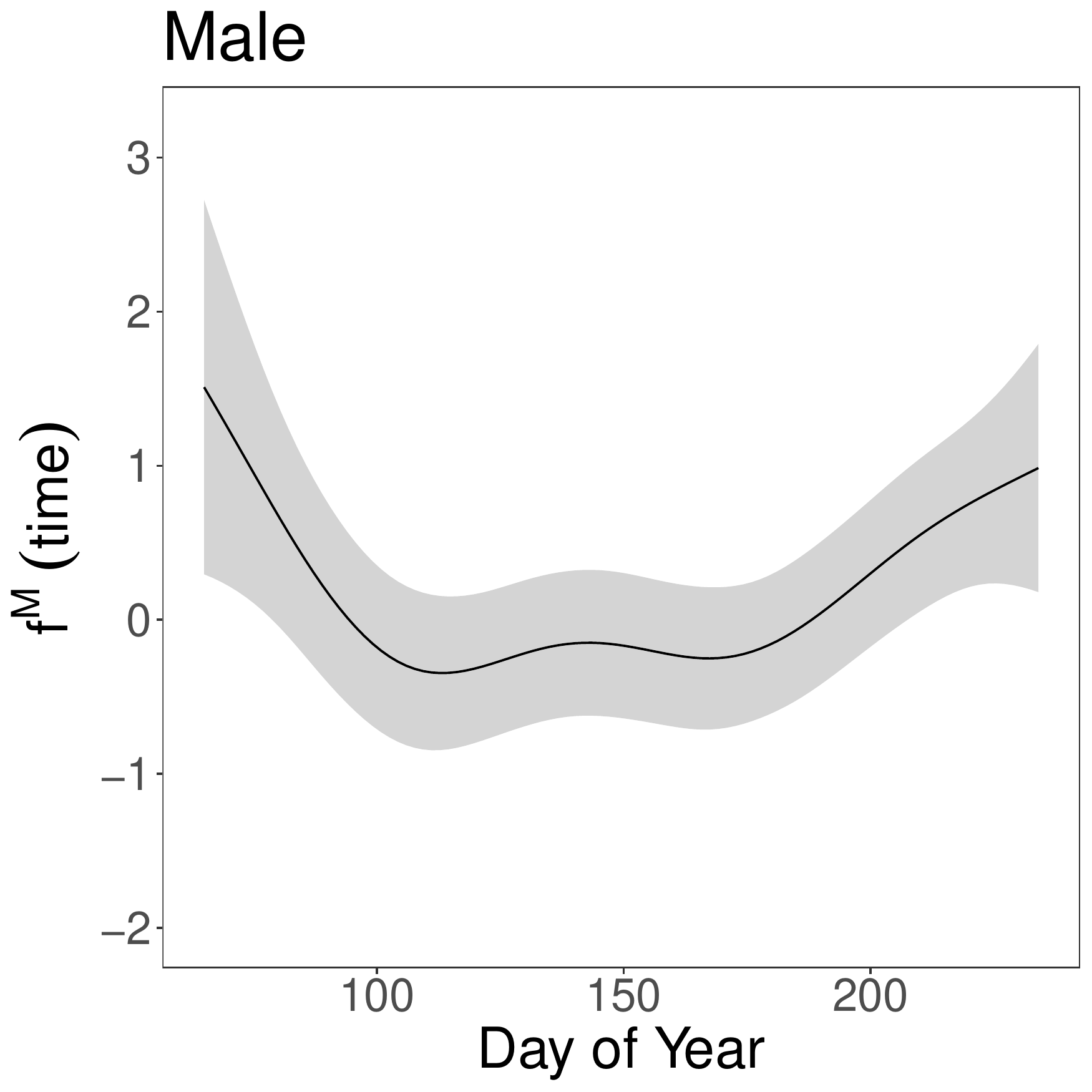}}
	\caption{Estimates of population-averaged seasonal trends in (a) female and (b) male beaver foraging behaviour (see \S\ref{ss:app2})}.

	\label{fig:bvfig}
\end{figure}

Estimated seasonal trends are plotted in Figure \ref{fig:bvfig}. While there is some evidence of increased foraging in trees/shrubs among males toward the end of the year, there was relatively little evidence of seasonality among male foraging behaviour. By contrast, we observed strong seasonal trends in population-averaged foraging behaviour among females. In particular, females tended to forage less in April and July, and comparatively more in May and June.

In addition to investigating seasonal trends, the MAM framework allowed us to characterize the dependence structure of the data. Interestingly, we observed more within-subject variability among females than among males even after accounting for differences in seasonal trends: the standard deviation of random intercepts was estimated to be 2.1 (95\% CI: [1.3, 3.4]) among females but only 0.7 (95\% CI: [0.4, 1.1]) among males.


\subsection{Spatial Analysis of Loa Loa Infection in West Africa}
\label{ss:app3}
Loa loa (``eye worm'') infection is an important public health risk in West Africa \citep{diggle2007spatial}. Data on the infection status of $\numunits=\sum_{i=1}^{\numclusters} \numunits_i = 25,771$ people in $\numclusters=190$ villages across Cameroon and Nigeria have been analyzed by \cite{loaloa} and \cite{prevmap}, and are available in the {\texttt{R}} package {\texttt{geostatsp}} \citep{geostatsp}. Spatially resolved covariates $\covx(\loc_i) = (x_{1}(\loc_i),x_{2}(\loc_i))$ are observed for each village having location $\loc_i\in\region\subset\R^{2}$ and include elevation and an index of vegetation. Interest lies in  quantifying the association between risk of infection and environmental factors like elevation and vegetation, as well as in characterizing the spatial variation in disease across the study area, and investigating between-village heterogeneity. 

\cite{loaloa,geostatsp,prevmap} analyzed these data as binomial counts $\response=(\responsei_1,\ldots,\responsei_\numclusters), \responsei_i\in N$, with known maxima $0\leq \responsei_i \leq \numunits_i , i\in[\numclusters]$ equal to number of people tested  in the $i^{th}$ village.  
While all allowed for spatial autocorrelation \textit{between} villages, the binomial likelihood assumed that people's disease statuses were independent \textit{within} the same village. A cursory fit of a ``quasi-binomial'' model yields an estimated dispersion parameter of $7.13\gg 1$, suggesting this assumption of within-village independence, and hence Binomially-distributed counts, may be inappropriate.

Instead, we explicitly model within-village dependence via a random intercepts structure, treating individual test results as correlated binary responses: for $i\in[\numclusters],j\in[\numunits_i]$, let $\responsei_{ij}\in\left\{0,1\right\}$ such that $\responsei_{i} = \sum_{j=1}^{\numunits_i}\responsei_{ij}$. Interest lies in associations of infection with elevation and vegetation, as well as spatial variation in infection risk, \textit{averaged over village locations}, so we adopt the following MAM:
%
\begin{equation}\label{eqn:loaloacond}
	\begin{aligned}
		\responsei_{ij} | \margmean\left\{\covx(\loc_i),\loc_i|\re_i\right\}&\ind\text{Bernoulli}\left[\margmean\left\{\covx(\loc_i),\loc_i|\re_i\right\}\right], \\
		\logit\left[ \margmean\left\{\covx(\loc_i),\loc_i\right\}\right] &= \margbetai_0 +  \smoothfunM_1\{x_{1}(\loc_i)\} + \smoothfunM_2\{x_{2}(\loc_i)\} + \smoothfunM_s(\loc_i), \\
		\logit\left[ \condmean\left\{\covx(\loc_i),\loc_i|\rei_i\right\}\right] &= \Delta(\loc_i)+ \rei_i, ~~	\rei_i\ind\text{Normal}\left(0,\varparamind^{2}\right), \\
		\text{Cov}\left\{ \smoothfunM_s(\loc_i),\smoothfunM_s(\loc_j)\right\}&=\varparam^{2}M(\norm{\loc_i - \loc_j}/\rangeparam),
	\end{aligned}
\end{equation}
where $M(\cdot)$ is a Matern correlation function \citep[Appendix A]{geostatsp}. As a byproduct of the random intercepts structure, we can characterize  between-village heterogeneity (or, equivalently, within-village dependence) via $\varparamind$.  In addition to the MAM  in (\ref{eqn:loaloacond}), we also fit an analogous binomial GAM that assumes within-village independence (fixing $\sigma_v=0$).

\begin{figure}[htbp!]
	\centering
	\subfigure[$\smoothfunMest(\texttt{evi})$, Independent GAM ]{\includegraphics[width=.49\textwidth]{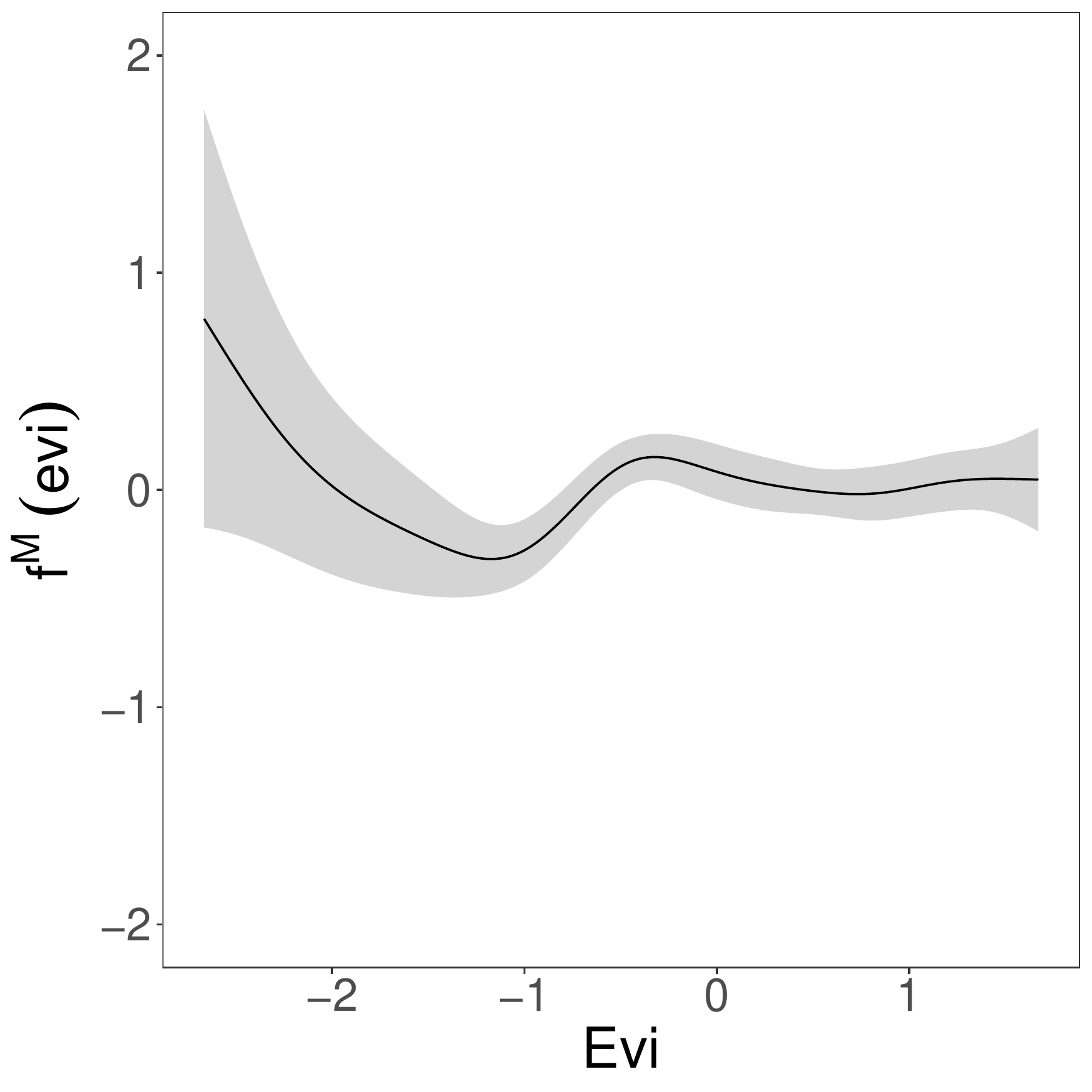}}
	\subfigure[$\smoothfunMest(\texttt{evi})$, MAM  (\ref{eqn:loaloacond}) ]{\includegraphics[width=.49\textwidth]{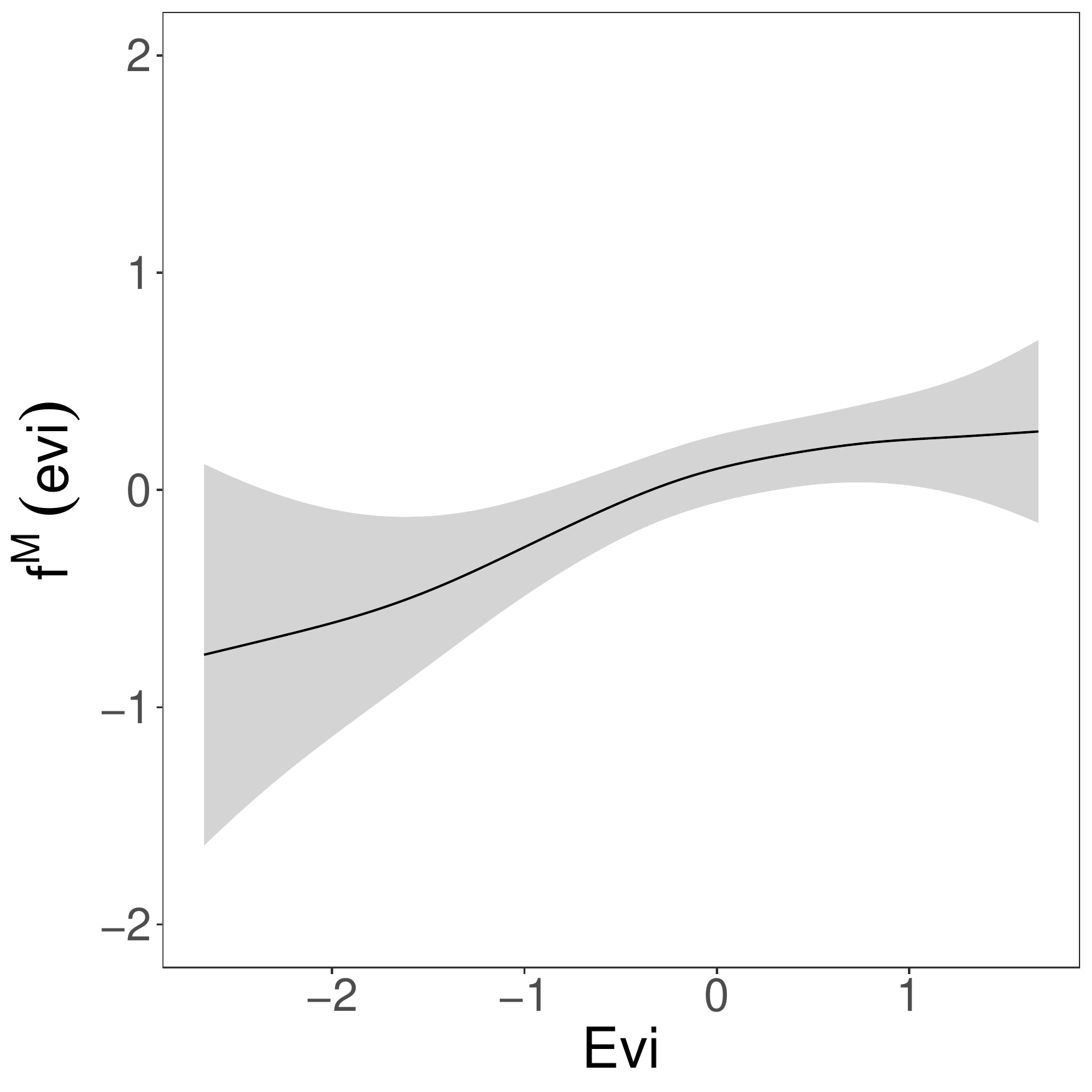}}
	
	\subfigure[$\smoothfunMest(\texttt{elevation})$, Independent GAM ]{\includegraphics[width=.49\textwidth]{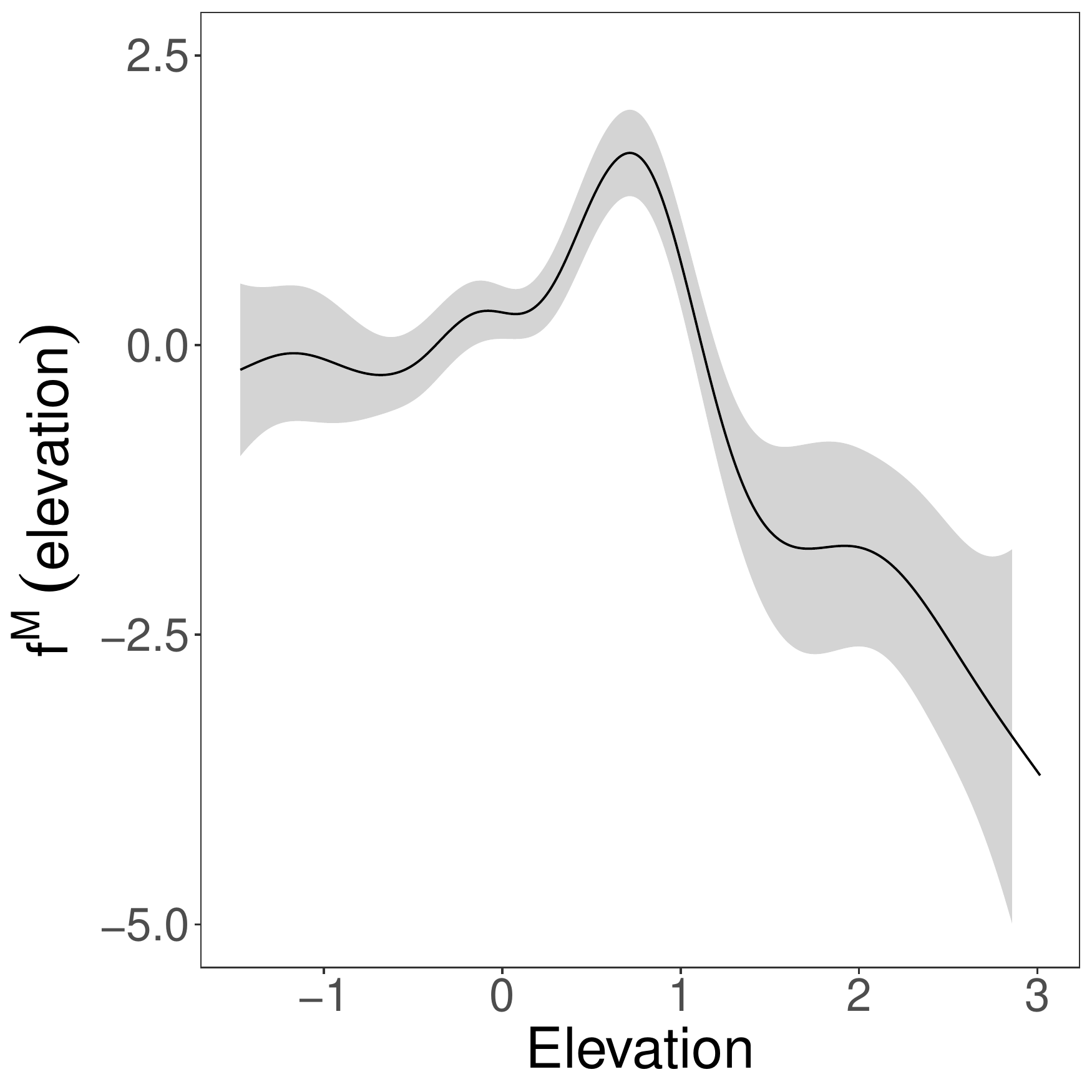}}
	\subfigure[$\smoothfunMest(\texttt{elevation})$, MAM  (\ref{eqn:loaloacond}) ]{\includegraphics[width=.49\textwidth]{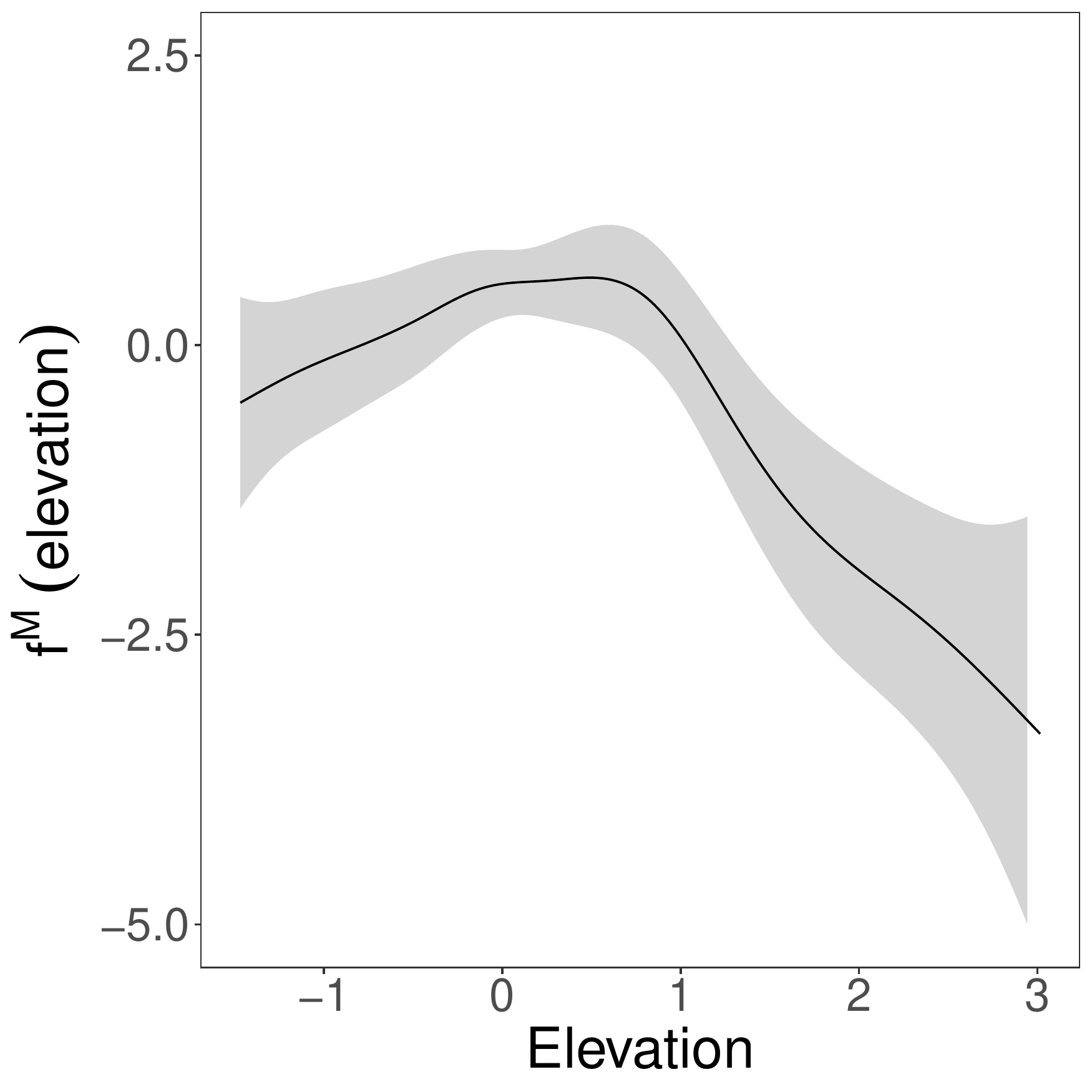}}
	
	\caption{Smooth covariate associations for vegetation index \texttt{evi} (a,b) and \texttt{elevation} (c,d) for the independent Binomial (a,c) and dependent Bernoulli (b,d) models.}
	\label{fig:loaloasmooth}
\end{figure}

Figure \ref{fig:loaloasmooth} shows estimated associations with the vegetation index (panels a \& b) and elevation (panels c \& d). The MAM estimates are generally attenuated and more uncertain relative to those of the GAM assuming independence, but still suggest strong-linearity.  Comparing the MAM to the independent GAM, the estimated associations with vegetation differ particularly strongly in the left tails, although these correspond to very large intervals.

\begin{figure}[htbp!]
	\centering
	\subfigure[Lower $95\%$ CI]{\includegraphics[width=.49\textwidth]{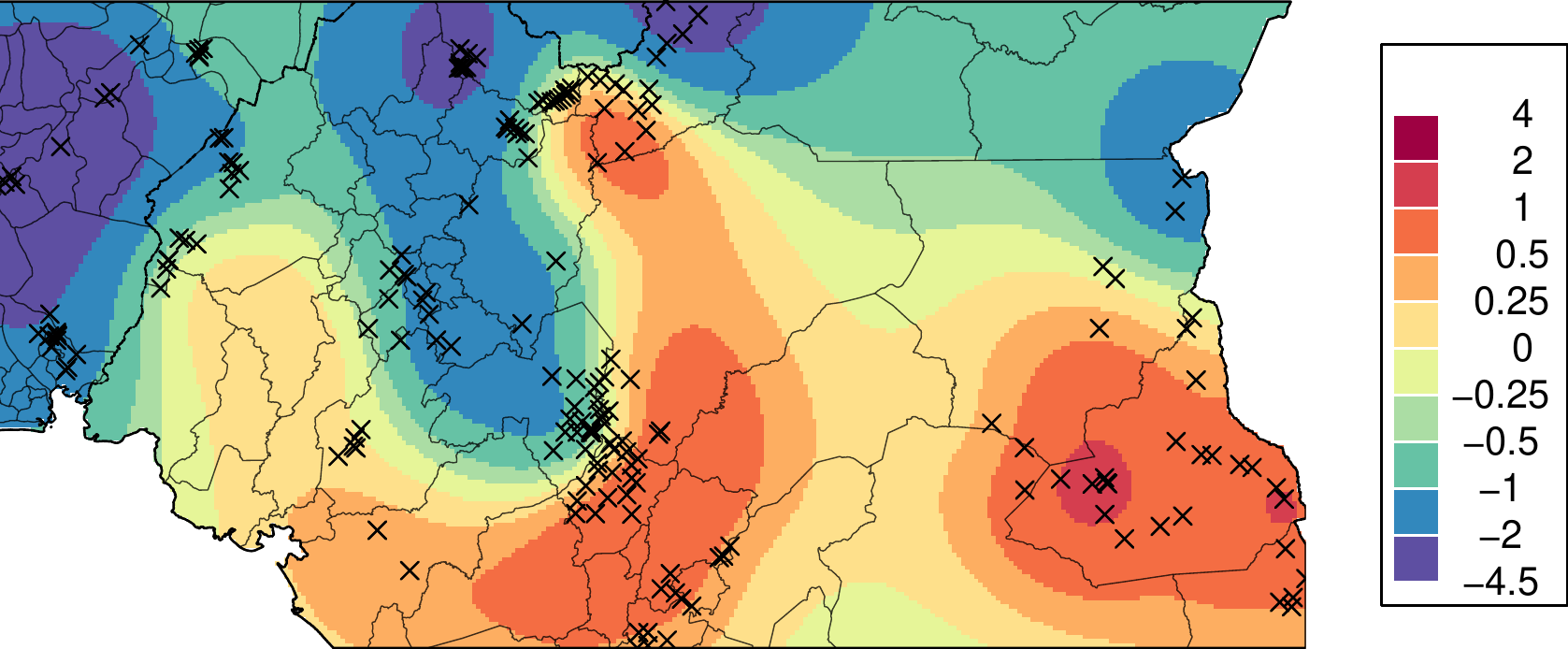}}
	\subfigure[Lower $95\%$ CI]{\includegraphics[width=.49\textwidth]{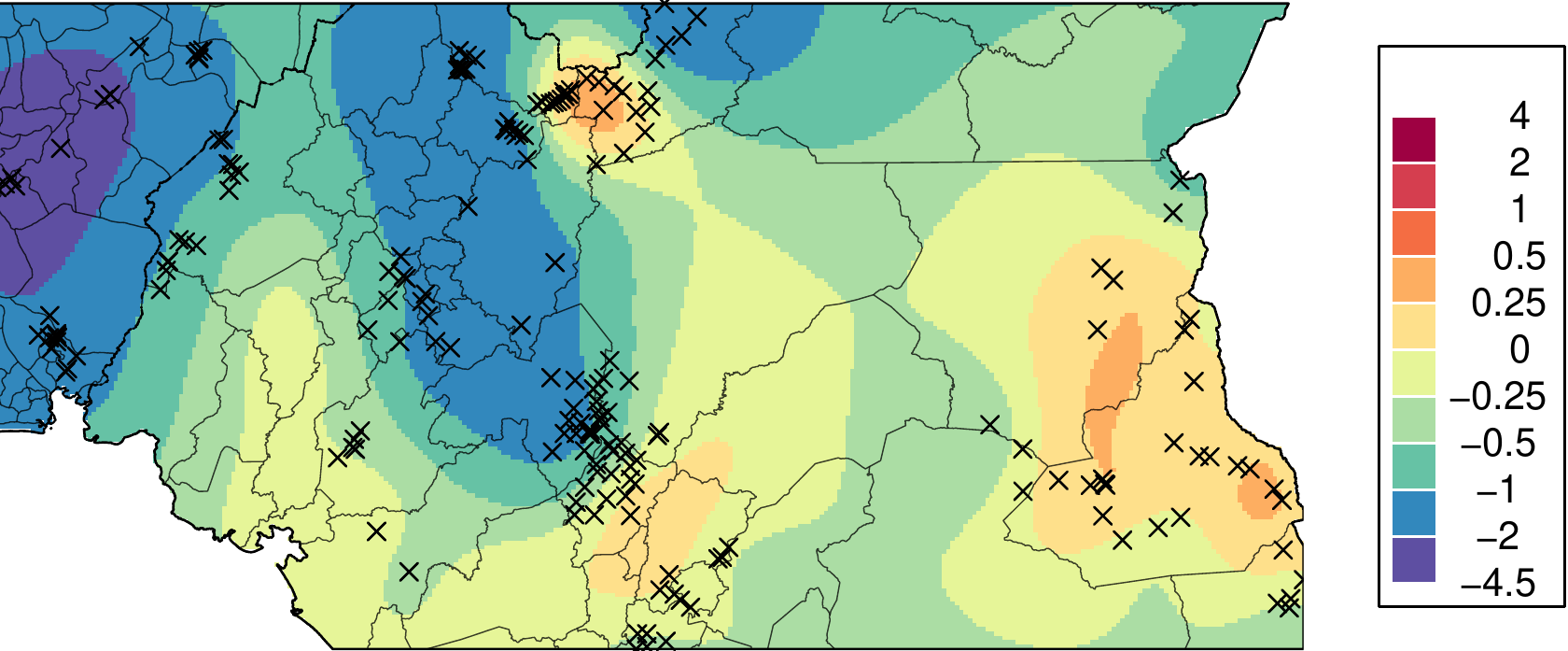}}
	
	\subfigure[Independent GAM , $\smoothfunM(\loc)$]{\includegraphics[width=.49\textwidth]{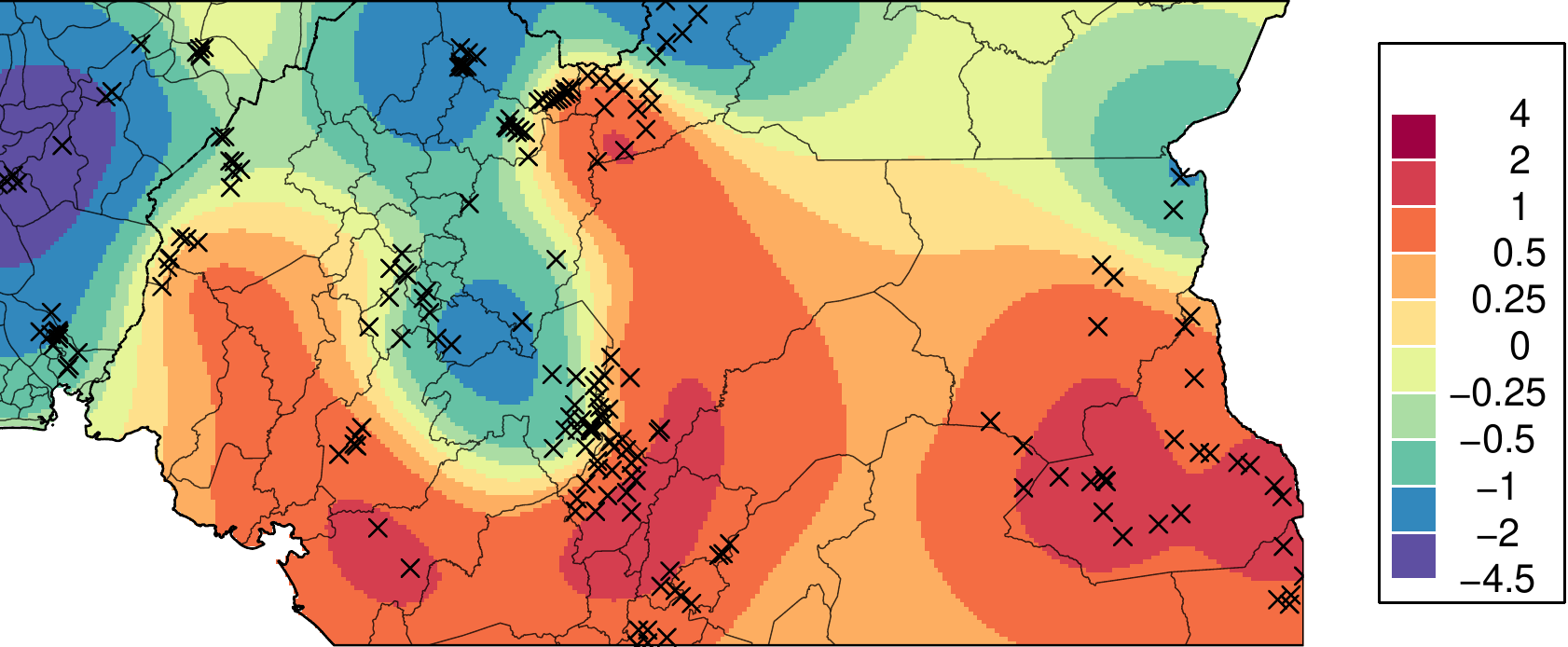}}
	\subfigure[MAM (\ref{eqn:loaloacond}), $\smoothfunM(\loc)$]{\includegraphics[width=.49\textwidth]{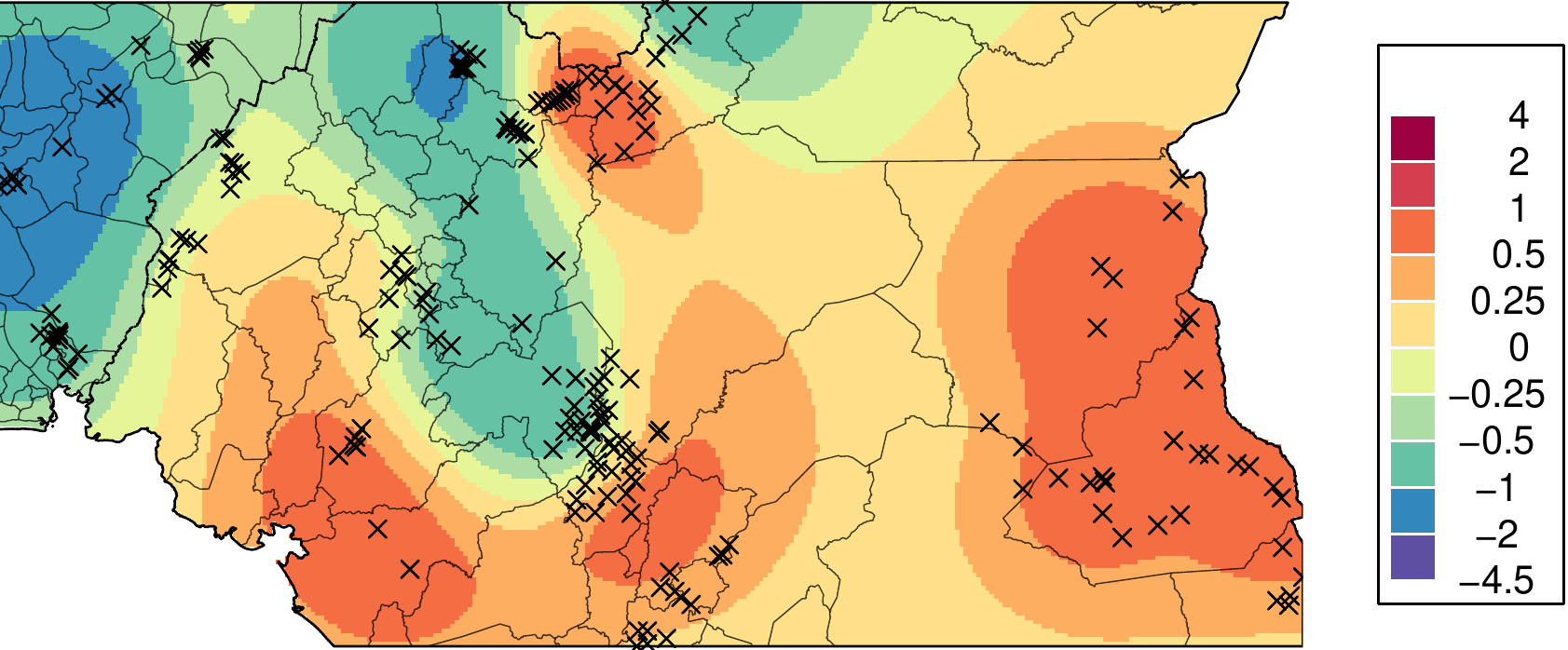}}
	
	\subfigure[Upper $95\%$ CI]{\includegraphics[width=.49\textwidth]{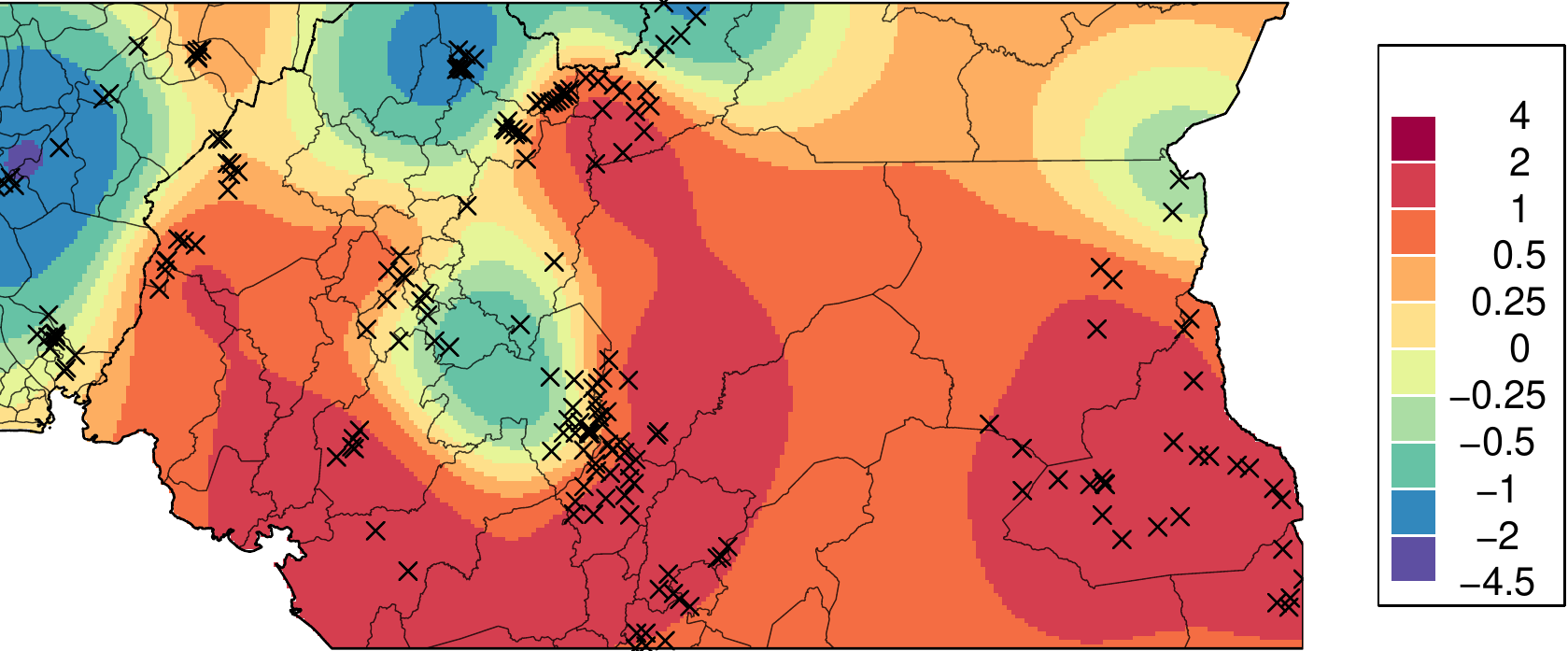}}
	\subfigure[Upper $95\%$ CI]{\includegraphics[width=.49\textwidth]{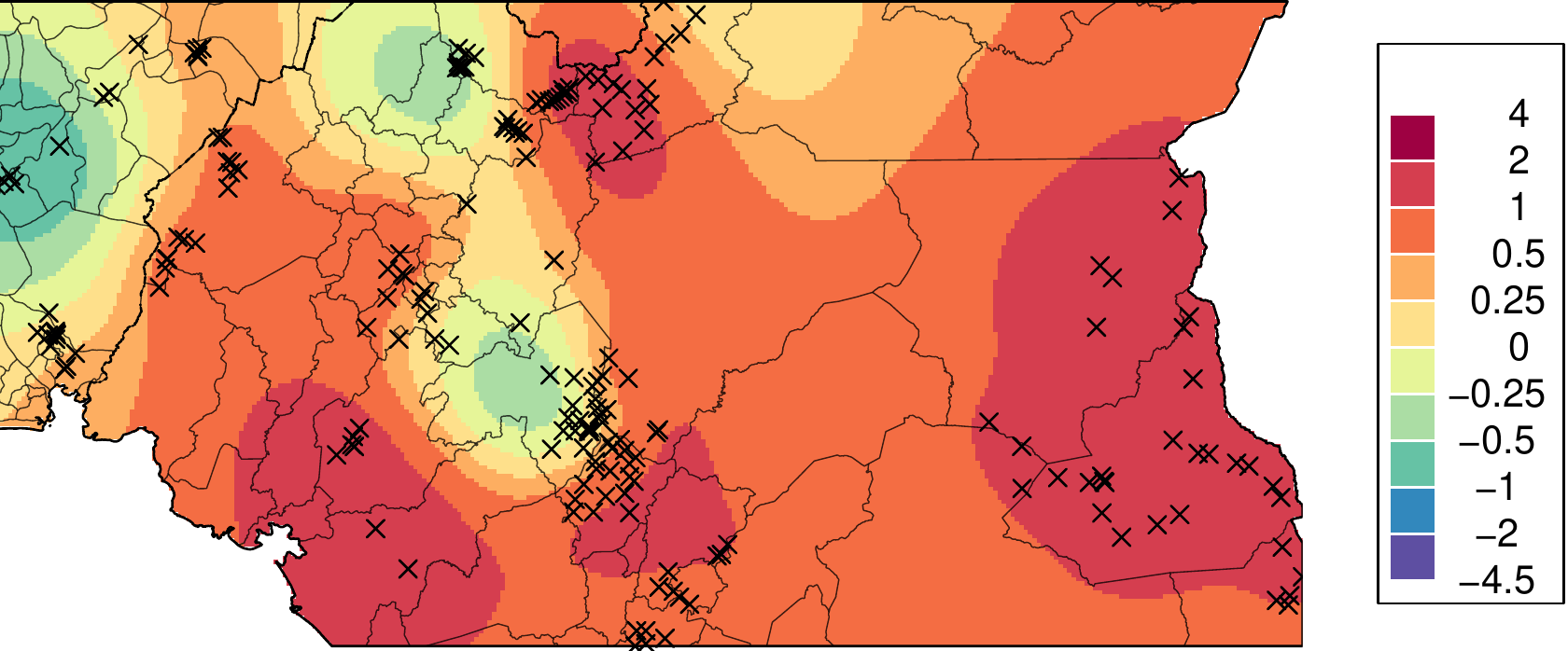}}
	
	\hfill	\subfigure[Spatial distribution of $\reest$]{\includegraphics[width=.49\textwidth]{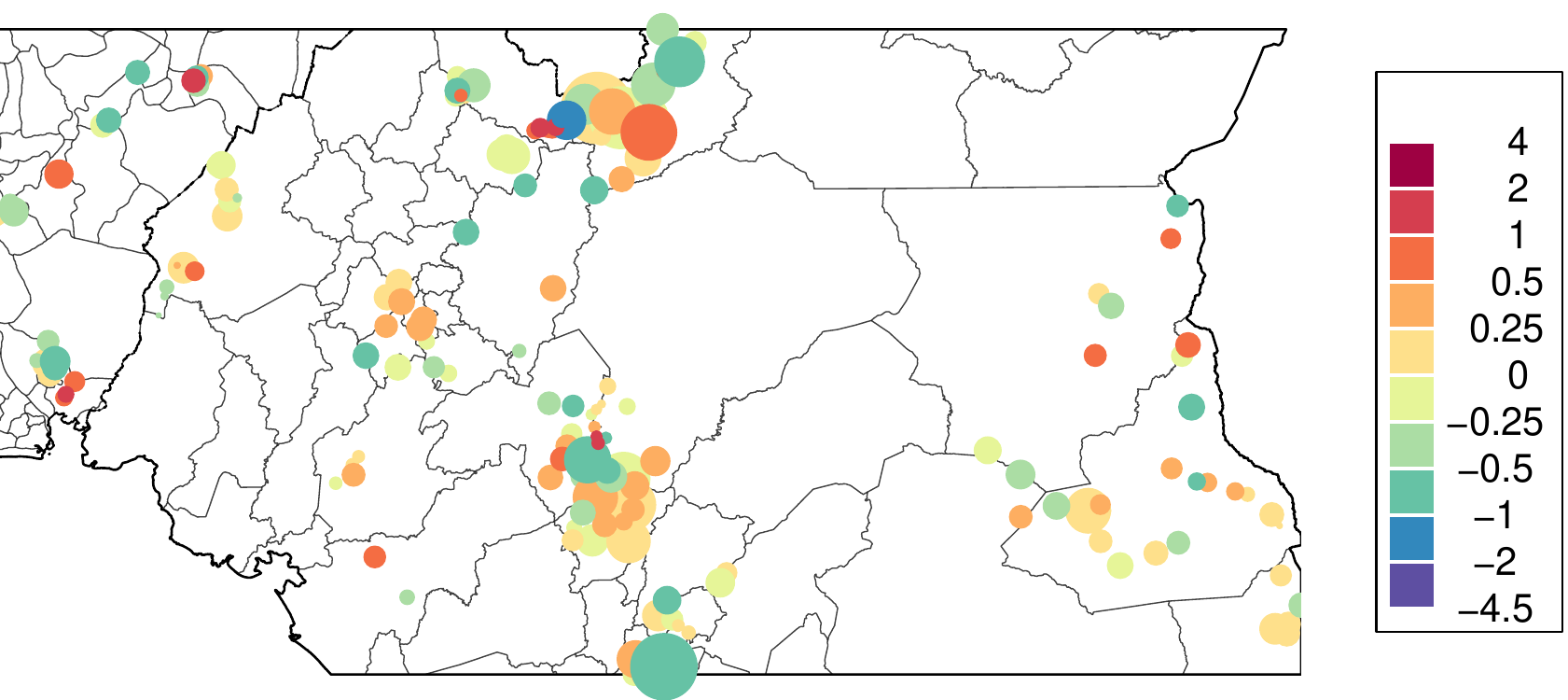}}

	\caption{Results of the Loa loa analysis under the independent binomial GAM (left column) and the proposed MAM  that models within-village dependence (right column). Estimates of the spatial effect $\smoothfunM_s(\loc_i)$ are shown in  the second row (panels c \& d), along with pointwise lower (first row; panels a \& b) and upper (third row; panels e \& f) $95\%$ confidence limits. Panel (g) shows village-specific random intercepts $\reest$; colour corresponds to value and size to number of residents.
	}
	\label{fig:loaloaspatial}
\end{figure}

Figure \ref{fig:loaloaspatial} shows estimates and approximate pointwise $95\%$ confidence maps for $\margmean\left\{\covx(\loc),\loc\right\}$. 
The map estimated by the MAM is more smoothly varying and attenuated than that of the independence model, which attributes all residual variation to spatial dependence.  

The estimated standard deviation of the village-level intercepts was $\varparamindest = 0.61$, with approximate $95\%$ confidence interval $(0.51,0.72)$. This corresponds to a median (adjusted) odds ratio  across pairs of villages, representing a ``village effect'' on the odds ratio scale, of 1.78 (1.62,1.99), suggesting substantial between-village heterogeneity \citep{larsen2000interpreting}. Panel (g) of Figure \ref{fig:loaloaspatial} shows predicted village intercepts $\reest$ and  highlights that even villages that are close in space have highly distinct village-level risks. For example, the adjacent blue and red points in the top middle of Figure \ref{fig:loaloaspatial} (g) have the two largest negative random effects ($<-1$) and two of the six largest positive effects ($>1$), respectively. Despite their proximity, these villages have very different estimated infection risk, an observation that could not have been made in the previous models that assume independence of observations within villages.

\section{Discussion}
\label{s:disc}	

The Marginal Additive Model framework estimates non-linear marginal associations.
We presented a longitudinal analysis of beaver behaviour in \S\ref{ss:app2} and a spatial analysis of a tropical disease in \S\ref{ss:app3}. In both applications, the objects of interest
were themselves non-linear, and hence fitting a Marginal Additive Model was an appropriate strategy. 
Nevertheless, the proposed methods are also  useful when scientific interest lies in simpler 
associations, like the effect of a categorical treatment.
Common confounders like age often have non-linear effects on health outcomes, and improperly modelling such factors can lead to biased estimates of treatment effects \citep{benedetti2004using}. The Marginal Additive Model can estimate a marginal treatment effect while controlling for non-linear confounding or even non-linear effect modification.

There has been much disagreement about if and when the conditional or marginal frameworks are appropriate
(\citealp{pendergast1996survey,neuhaus1991comparison,graubard1994regression,lindsey1998appropriateness,zeger1988models,lee2004conditional}). \citet{lee2004conditional} adopt the perspective that the conditional model is fundamental, and that marginal predictions can be made from it if required. 
In contrast, \citet[ \S8.1]{digglelongitudinal} argue that marginal models should be fit in cases when the marginal associations themselves are of primary inferential interest. (See the discussion to \citealp{lee2004conditional} for further elaboration.) A key contribution of the Marginal Additive Model is that it estimates both conditional and marginal associations simultaneously, inheriting the advantages of both frameworks, and allowing the analyst to mix-and-match inferences about clusters and the population of interest within the same analysis.

\section*{Software \& Data Availability}
Software for implementing the methodology considered here is available at \url{https://github.com/awstringer1/mam}. Code for running the simulations in Section 4 and the analyses in Section 5 is available at \url{https://github.com/awstringer1/mam-paper-code}. Data used in Section 5 are openly available in the referenced \texttt{R} packages.


%
%

	\bibliographystyle{chicago} 
	\bibliography{Bibliography}


\clearpage
\begin{centering}

\section*{Supplementary Appendices}
\end{centering}

\appendix
\counterwithin{figure}{section}
%
%
%
%
%
%
%
%
%
%
%
%
%
%

\section{Algorithms}

\listofalgorithms
\addcontentsline{toc}{section}{List of algorithms}

\begin{algorithm}[p]
	\textbf{Input}:
	\begin{itemize}
		\item[] $\numclusters\in\N$, $\numunits_i\in\N,i\in[\numclusters]$, $n = \sum_{i=1}^{\numclusters}\numunits_i$, $\smoothdim\in\N$, $\Udim\in\N$, $\quadnum\in\N$.
		\item[] $\data = (\datai_{ij})_{i\in[\numclusters],j\in[\numunits_{i}]}\in\R^{n}$, $\covx_{ij}\in\R^{\covxdim},\covz_{ij}\in\R^{\Udim},i\in[\numclusters],j\in[\numunits_{i}]$.
		\item[] $\responsedens(\cdot):\R^{n}\to\R^{+}$, $\redens(\cdot):\R^{\Udim}:\R^{+}$, 
		\item[] $\left\{\penaltyP_{\penaltyC_{l}}(\condbasiscoef_{l};\penaltymat_{l})\right\}l\in[\smoothdim]$, $\left\{\condbasisfun_{lq}(\cdot)\right\}_{l\in[\smoothdim],q\in[\basisdim_l]}$, $\left\{\margbasisfun_{lq}(\cdot)\right\}_{l\in[\smoothdim],q\in[\basisdim_l]}$,
		\item[] $\quadpointset(\Udim,\quadnum)\subset\R^{\Udim}$, $\weight_{\quadnum}:\quadpointset(\Udim,\quadnum)\to\R^{+}$
	\end{itemize}
	\textbf{Transform}:
	\begin{itemize}
		\item[] $\link\left\{\condmeani(\covx_{ij},\covz_{ij}|\re_{i})\right\}\ =\ \sum_{l=1}^p\smoothfunC_l(\covxi_{ijl})+ \covz_{ij}\tpose \re_{i}$, $\smoothfunC_l(x)=\sum_{q=1}^{\basisdim_l} \condbasisfun_{lq} (x) \condbasiscoefi_{lq}$,
		\item[] $\condXbasis\in\R^{n\times \sum_l^p Q_l}$; $\condXbasis_{ij}=[{{\condbasisfunvec}_{ij1}}\tpose,\dots,{\condbasisfunvec_{ijp}}\tpose]$; $\condbasisfunvec_{ijl}=[\condbasisfun_{l1} (\covxi_{ijl}),\dots,\condbasisfun_{lQ_l} (\covxi_{ijl}) ]\tpose$
		\item[] $\margXbasis\in\R^{n\times \sum_l^p Q_l}$; $\margXbasis_{ij}=[{{\margbasisfunvec}_{ij1}}\tpose,\dots,{\margbasisfunvec_{ijp}}\tpose]$; $\margbasisfunvec_{ijl}=[\margbasisfun_{l1} (\covxi_{ijl}),\dots,\margbasisfun_{lQ_l} (\covxi_{ijl}) ]\tpose$
		\item[] $\lik_{ij}(\condbasiscoef,\re_{i};\datai_{ij}) = \responsedens\left\{\condmeani(\covx_{ij},\covz_{ij}|\re_{i})\right\}\redens(\re_{i})$,
		\item[] $\penaltyP_{\penaltyCvec}(\condbasiscoef;\penaltymat_{1}\,\ldots,\penaltymat_{\smoothdim})=\prod_{l=1}^{\smoothdim}\penaltyP_{\penaltyC_{l}}(\condbasiscoef_{l};\penaltymat_{l})$,
		\item[] $\lik(\condbasiscoef,\re,\penaltyCvec,\covmat;\data) = \left\{\prod_{i=1}^{\numclusters}\prod_{j=1}^{n_j}\lik_{ij}(\condbasiscoef,\re_{i};\datai_{ij})\right\}\times\penaltyP_{\penaltyCvec}(\condbasiscoef;\penaltymat_{1}\,\ldots,\penaltymat_{\smoothdim})$,
		\item[] $\mlik(\penaltyCvec,\covmat;\data) = \int\lik(\condbasiscoef,\re,\penaltyCvec,\covmat;\data)d\redist(\re,\covmat)d\condbasiscoef$,
		\item[] $\mlikLA(\penaltyCvec,\covmat;\data)$, Laplace approximation to $\lik(\condbasiscoef,\re,\penaltyCvec,\covmat;\data)$.
	\end{itemize}
	\caption{Preparing to fit a Marginal Additive Model}
	\label{alg:prepare}
\end{algorithm}

\begin{algorithm}[p]
	\textbf{Input}: output of Algorithm \ref{alg:prepare}.
	
	\textbf{Compute}:
	\begin{enumerate}
		\item \emph{Conditional model}:
		\begin{itemize}
			\item[] $(\penaltyCestvec,\covmatest) = \text{argmax} \ \mlikLA(\penaltyCvec,\covmat;\data)$,
			\item[] $(\condbasiscoefest,\reest) = \text{argmax} \ \lik(\condbasiscoef,\re,\penaltyCestvec,\covmatest;\data)$,
			\item[] $\smoothfunCest(\covx_{ij}) = {\condXbasis_{ij}} \condbasiscoefest,i\in[\numclusters],j\in[\numunits_i]$.
		\end{itemize}
		\item \emph{Marginal means}:
		\begin{itemize}
			\item[] $\margmeanlinkesti(\covx_{ij}) = \link\left[ \sum_{\quadpoint\in\quadpointset(\Udim,\quadnum)}\invlink\left\{\smoothfunCest(\covx_{ij}) + \covz_{ij}\tpose\quadpoint\right\}\gqkernel\left\{\quadpoint;\covmatest\right\}\weight_{\quadnum}(\quadpoint)\right],i\in[\numclusters],j\in[\numunits_i]$,
			\item[] $\margmeanlinkest(\Xmat) = \left(\margmeanlinkesti(\covx_{ij})\right)_{i\in[\numclusters],j\in[\numunits_i]}\in\R^{n}$,
			\item[] $\Dmat = \partial\margmeanlinkest(\Xmat)/\partial\left(\condbasiscoef,\re,\penaltyCvec,\covmat\right)\in\R^{n\times(\basisdim + \numclusters\Udim + \smoothdim + \paramsmalldim)}$
		\end{itemize}
		\item \emph{Marginal model}:
		\begin{itemize}
			\item[] $\margbasiscoefest=\left({\margXbasis}\tpose  \margXbasis \right)^{-1}{\margXbasis}\tpose  \margmeanlinkest$,
			\item[] $\smoothfunMest(\Xmat)= {\margXbasis} \margbasiscoefest$
			\item[] $\left(\VarMfest\right)_{ii}, \left(\VarMmargest\right)_{ii}, i\in[n]$ according to Algorithm \ref{alg:standarderrors}.
		\end{itemize}
	\end{enumerate}
	\textbf{Report}:
	\begin{enumerate}
		\item \emph{Point estimates} $\smoothfunMest(\Xmat)$ of $\smoothfunM(\Xmat)$,
		\item \emph{Pointwise $95\%$ confidence bands}: $(\smoothfunMest(\Xmat))_{i}\pm2\times\sqrt{\left(\VarMfest\right)_{ii}+\left(\VarMmargest\right)_{ii}},i\in[n]$.
	\end{enumerate}
	\caption{Fitting a Marginal Additive Model}
	\label{alg:fit}
\end{algorithm}

\begin{algorithm}[p]
	\textbf{Input}: output of Algorithm \ref{alg:fit}, steps (1) and (2).
	
	\textbf{Compute}:
	\begin{enumerate}
		\item $\PrecCestcond=-\partial^{2}_{\condbasiscoef,\re}\log\lik(\condbasiscoefest,\reest,\penaltyCestvec,\covmatest;\data)$, $\PrecCestmarg = -\partial^{2}_{\penaltyCvec,\covmat}\log\mlikLA(\penaltyCvec,\covmat;\data)$.
		\item For $\PrecCestall=\PrecCestcond$:
		\begin{enumerate}
			\item $\chol=\texttt{Matrix::Cholesky(}\PrecCestall\texttt{,perm=TRUE,LDL=TRUE)}$.
			\item $\chol\tposeinv\Dmat\tpose\texttt{ <- sqrt(solve(}\chol\texttt{,system="D")) \%*\%}$
			\begin{sloppypar}
				$\texttt{solve(}\chol\texttt{,solve(}\chol\texttt{,}\Dmat\tpose\texttt{,system = "P"),system = "L")}.$
			\end{sloppypar}
			\item $\Vfac\tpose = \left\{ \chol\tposeinv\Dmat\tpose\right\}\margXbasis\left({\margXbasis}\tpose  \margXbasis \right)\inv{\margXbasis}\tpose$
			\item $\left(\VarMfest\right)_{ii} = \texttt{colSums}\left\{\left(\Vfac\tpose\right)^{2}\right\}_{i},i\in[n]$.
		\end{enumerate}
		\item \textbf{Repeat} (a) --- (d) with $\PrecCestall=\PrecCestmarg$, obtaining $\left(\VarMmargest\right)_{ii},i\in[n]$.
	\end{enumerate}
	\textbf{Output}: $\text{Var}(\smoothfunM(\covx_{i})) \approx (\VarMfest)_{ii} + (\VarMmargest)_{ii}, i\in[n]$.
	
	\caption{Standard Errors in a Marginal Additive Model}
	\label{alg:standarderrors}
\end{algorithm}

\clearpage
\section{Additional Simulation Results}

\subsection{Random Slopes}
\begin{figure}[tbph!]
	\centering
	\subfigure[GAM, $x_{1}$]{\includegraphics[width=0.49\linewidth]{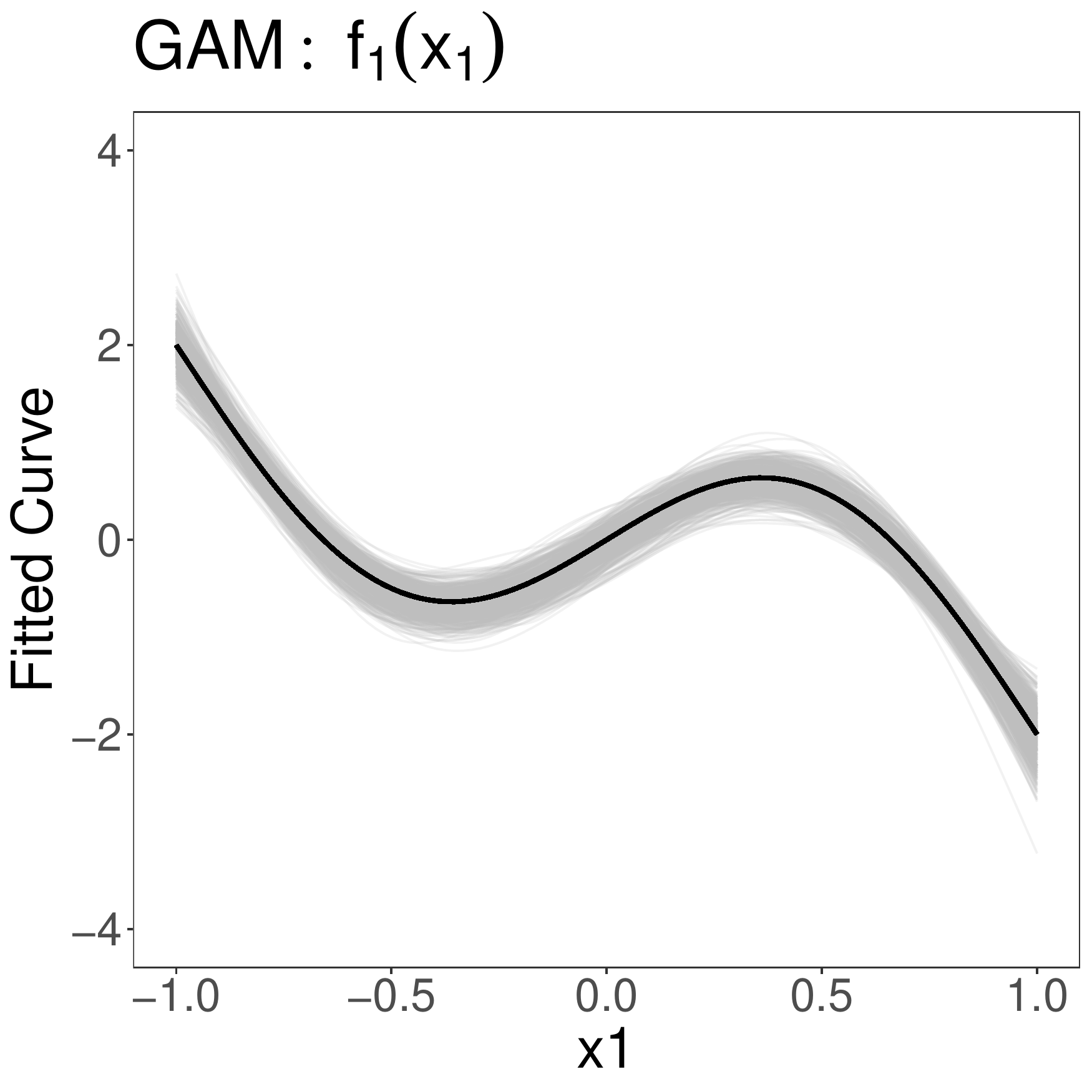}}
	\subfigure[MAM, $x_{1}$]{\includegraphics[width=0.49\linewidth]{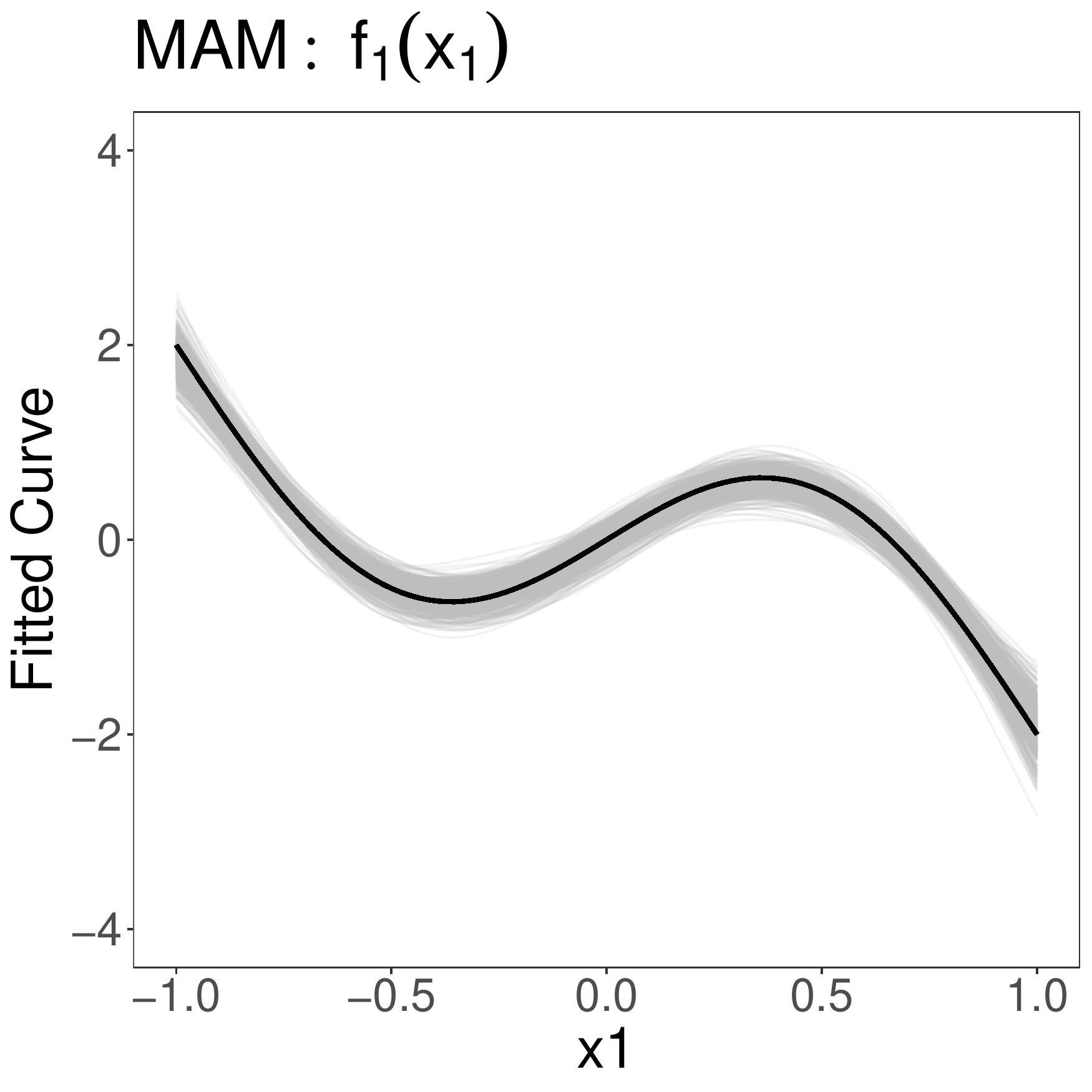}} \\

	\subfigure[GAM, $x_{2}$]{\includegraphics[width=0.49\linewidth]{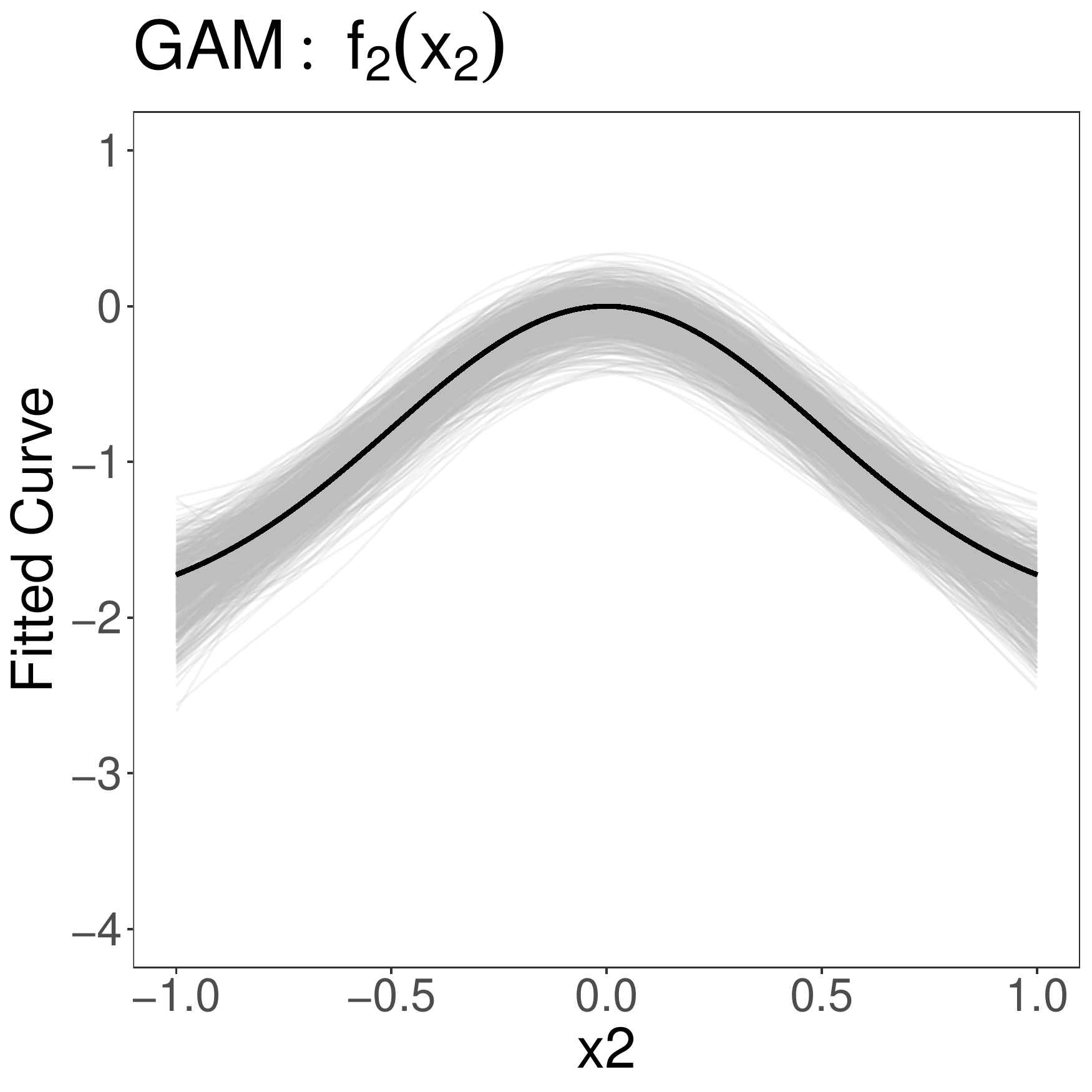}}
	\subfigure[MAM, $x_{2}$]{\includegraphics[width=0.49\linewidth]{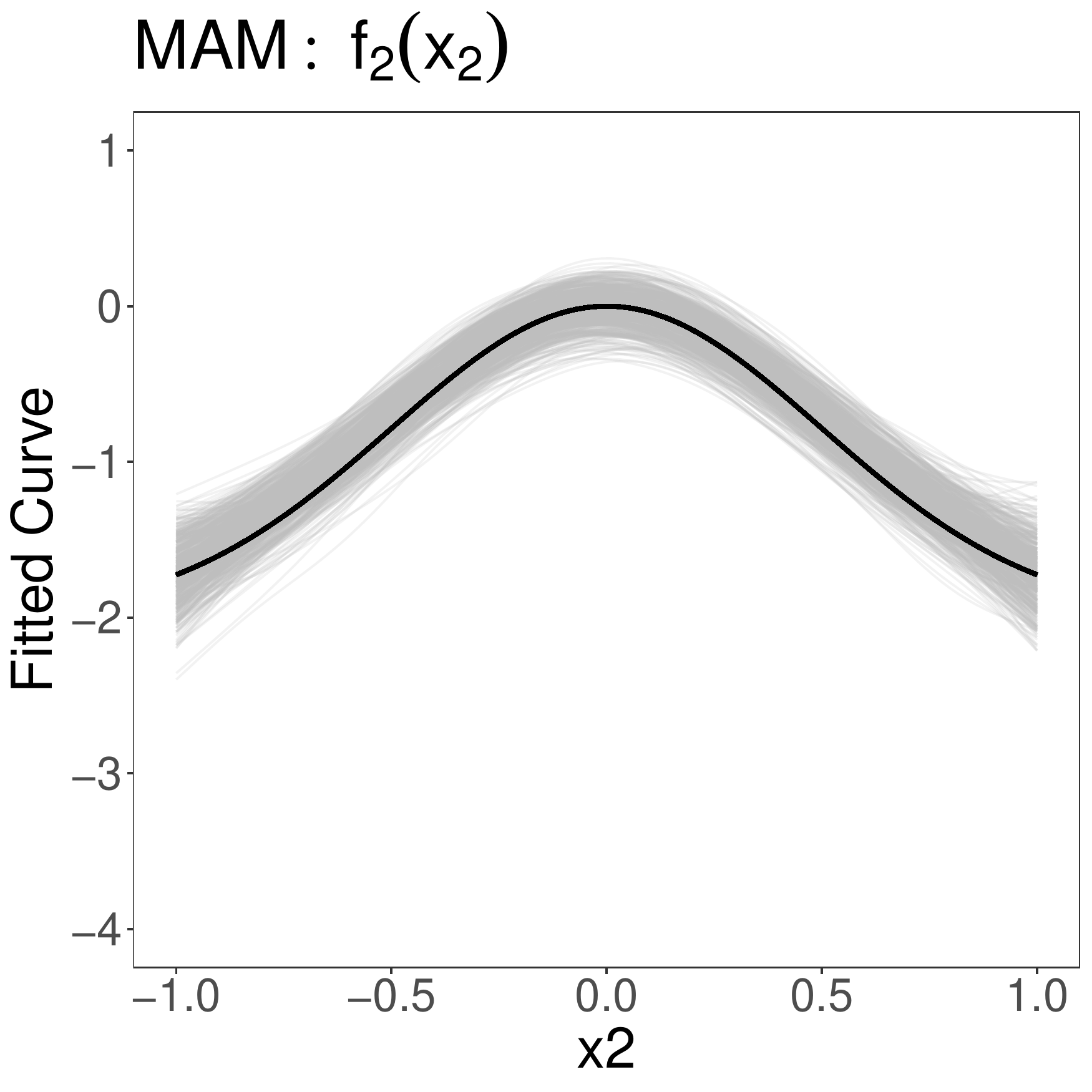}}
	\caption{Simulation results for marginal mean model under random slopes model with $\numclusters=200$ and $\numunits_{i}=20$. Figure shows distributions of estimates (top row) for  $f_1(\cdot)$ and $f_2(\cdot)$ on a grid of evenly spaced covariate values.}
	\label{fig:sim1slopes}
\end{figure}

\begin{figure}[tbph!]
	\centering
	\subfigure[GAM, $x_{1}$]{\includegraphics[width=0.49\linewidth]{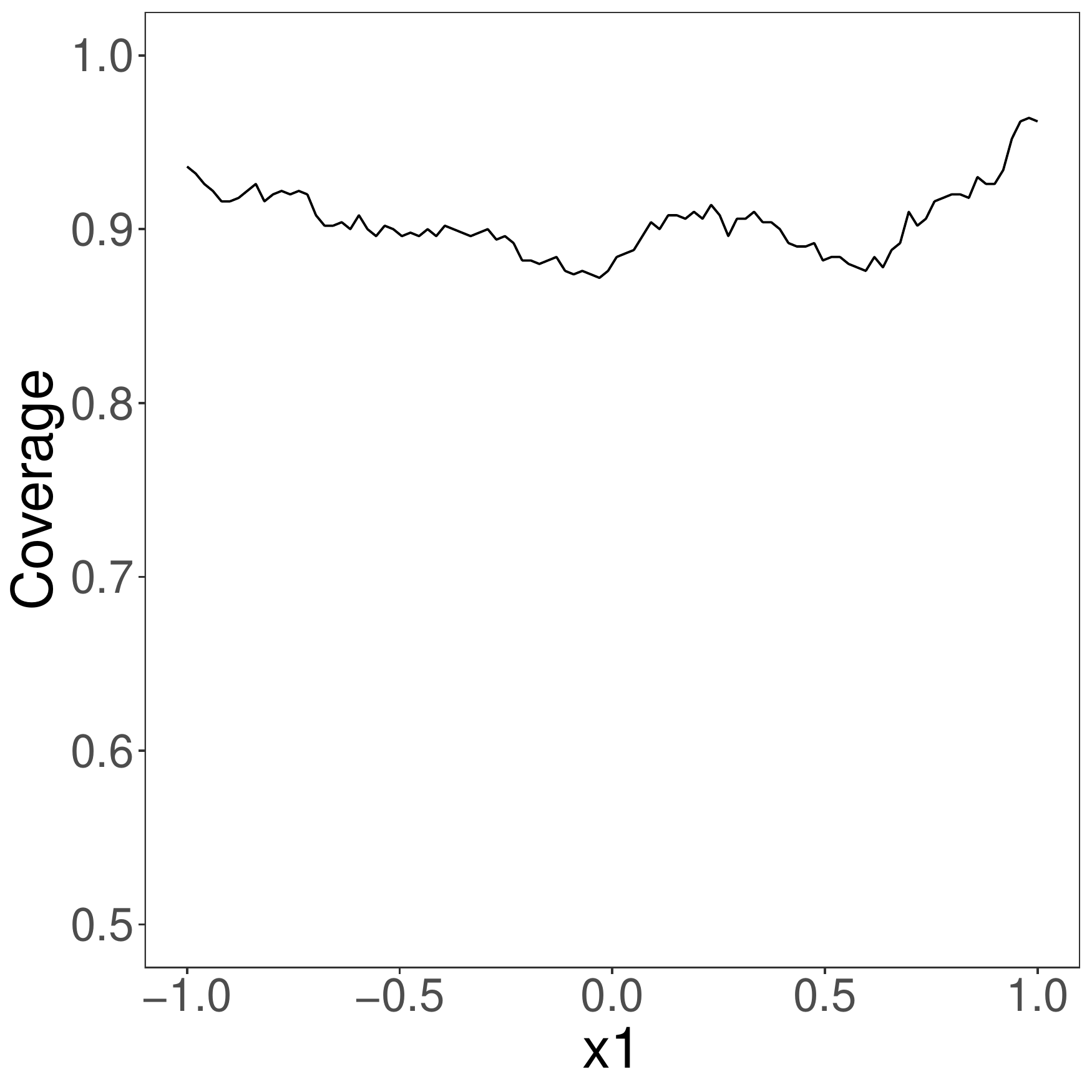}}
	\subfigure[MAM, $x_{1}$]{\includegraphics[width=0.49\linewidth]{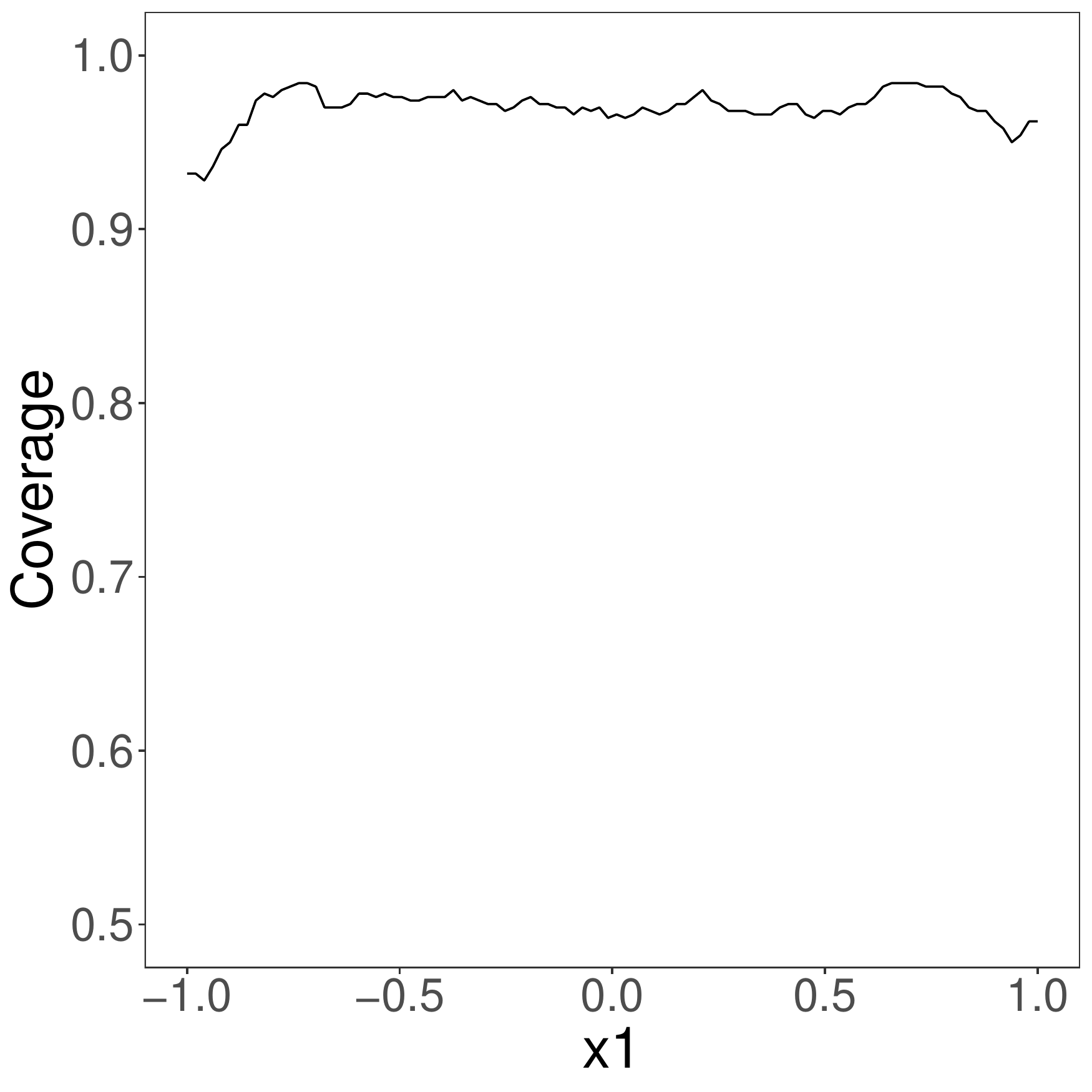}} \\

	\subfigure[GAM, $x_{2}$]{\includegraphics[width=0.49\linewidth]{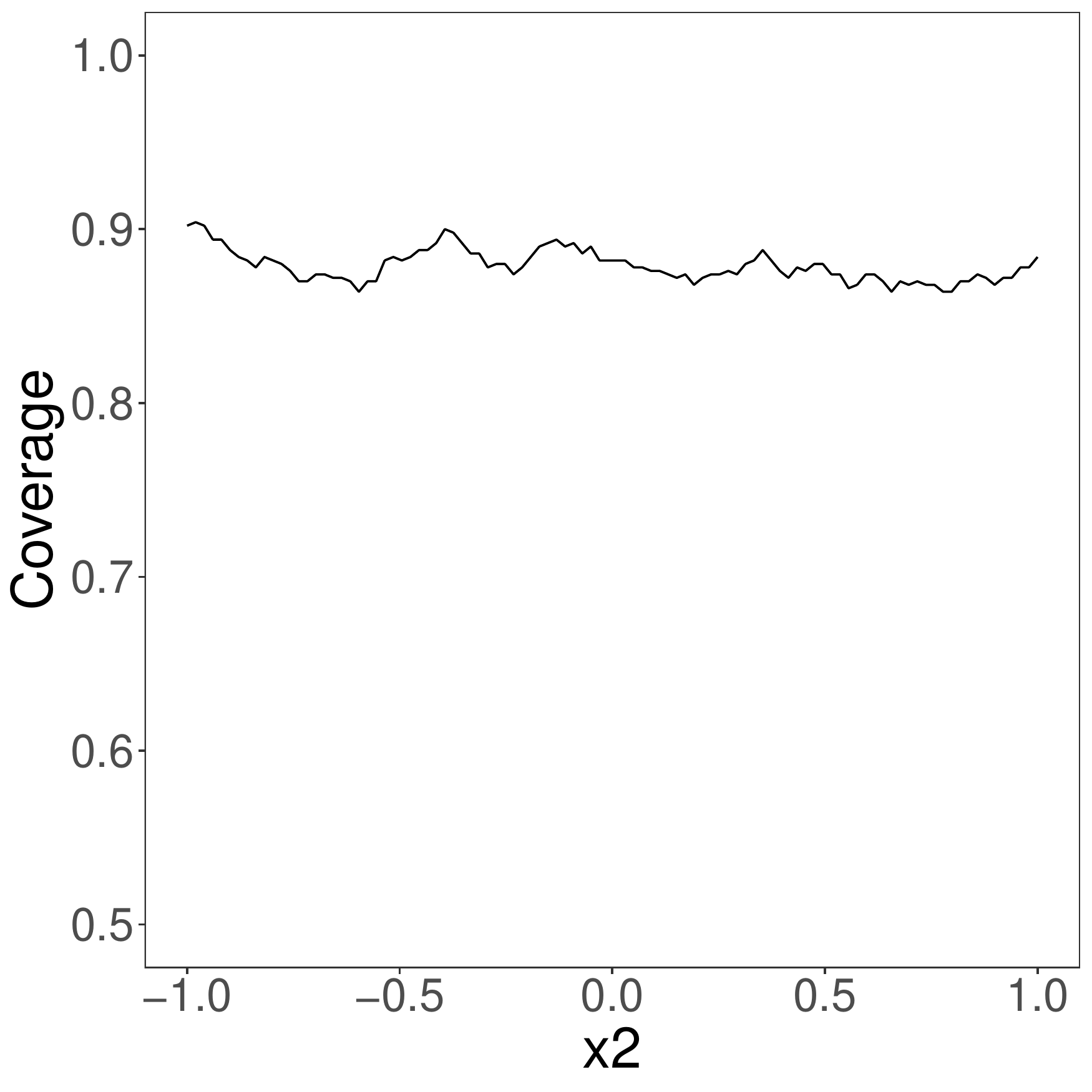}}
	\subfigure[MAM, $x_{2}$]{\includegraphics[width=0.49\linewidth]{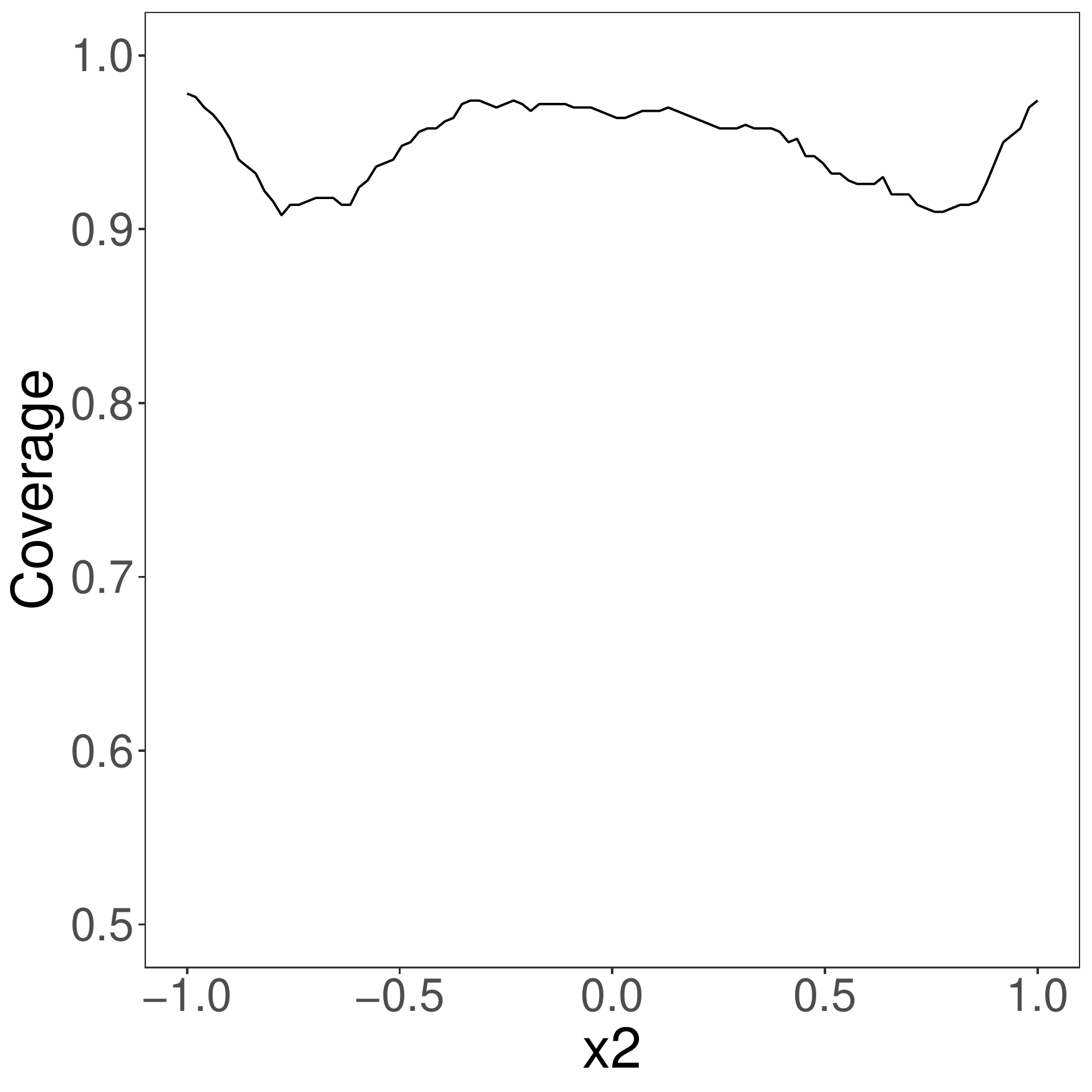}}
	\caption{Simulation results for marginal mean model under random slopes model with $\numclusters=200$ and $\numunits_{i}=20$. Figure shows confidence interval coverages for  $f_1(\cdot)$ and $f_2(\cdot)$ on a grid of evenly spaced covariate values.}
	\label{fig:sim1slopescovr}
\end{figure}

\begin{figure}[tbph!]
	\centering
	\includegraphics[width=1.0\linewidth]{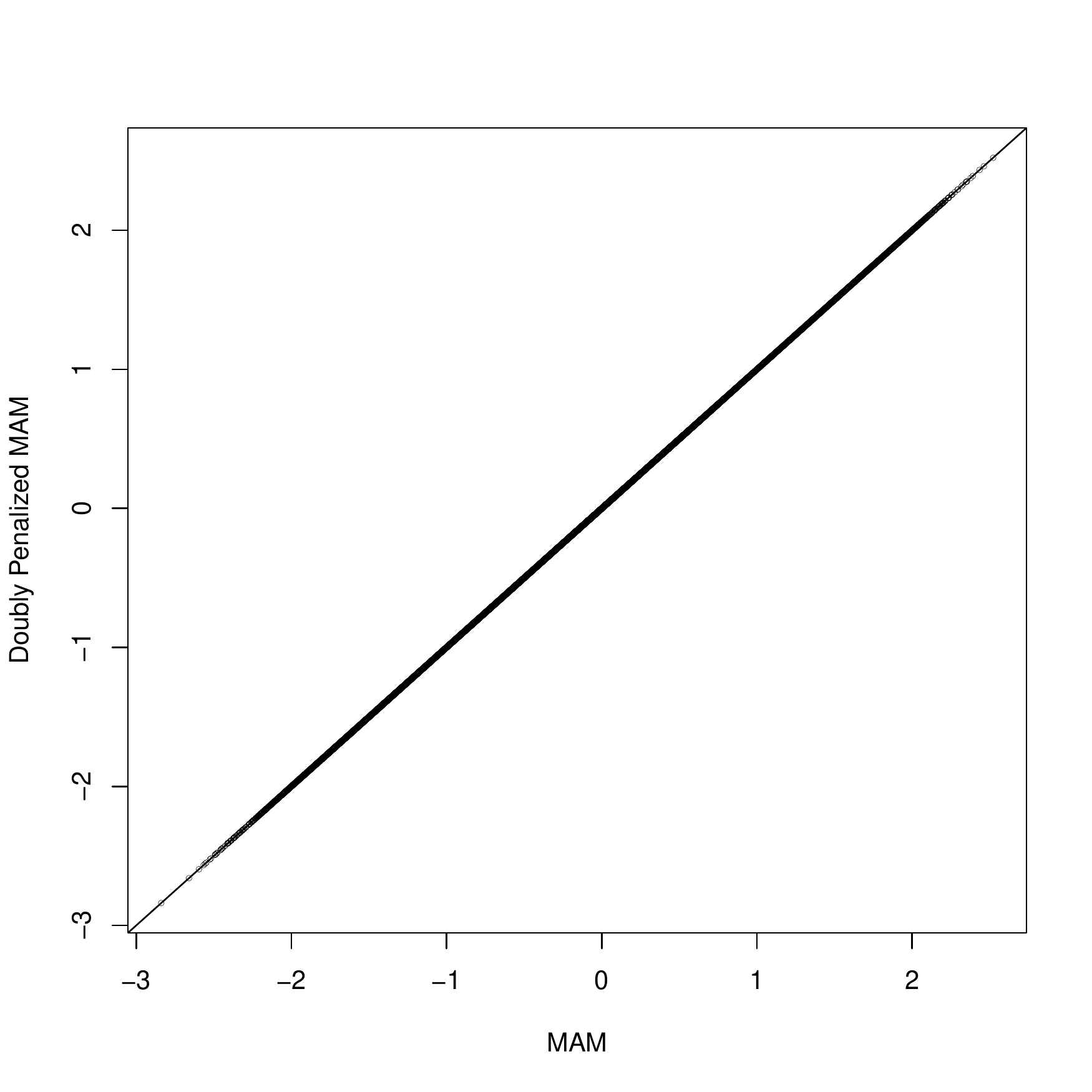}
	\caption{Simulation results for marginal mean model under random slopes with $\numclusters=200$ and $\numunits_{i}=20$. Figure shows fitted values across 500 datasets, comparing the proposed estimation strategy of \S3.2 with an alternative strategy in which marginal estimates are subject to a second penalization:  $\margbasiscoefest_{double}=\text{argmin}\norm{\margmeanlinkest - \margXbasis\margbasiscoef}  +\sum_{l=1}^p \penaltyP_{\penaltyM_{l}}(\margbasiscoef_{l};\penaltymat_{l})$ where $\penaltymat_{l}$ is a $\basisdim_{l}$-dimensional fixed, known penalty matrix with rank $\rank_{l}\leq \basisdim_{l}$ and $\penaltyM_{l}>0$ is a ``smoothing'' parameter to be estimated.}
	\label{fig:doublepenaltyslopes}
\end{figure}

\clearpage
\subsection{Random Intercepts Only}

\begin{table}[htbp!]
	\centering
	\caption{Simulation results. Bias and 95\% confidence interval coverage (Cvg; in \%) for estimates of $f_1(x_1)$ and $f_2(x_2)$ averaged over a grid of evenly spaced values of $x_1$ and $x_2$, bias in estimates of $\sigma_0$, root mean squared error of prediction (RMSEP) for random intercepts $\rei_{i0}$. $\numclusters\in\{100,200\},\numunits_i \in \{10,20\},\sigma_0\in\{1,2\}$ \label{tab:sim1} }
	\begin{tabular}{l l l l r@{\extracolsep{5pt}}r r@{\extracolsep{5pt}}r r@{\extracolsep{5pt}}@{\extracolsep{5pt}}r}
		\toprule
		&&& & \multicolumn{2}{c}{$\widehat{f_1}(x_1)$ } & \multicolumn{2}{c}{$\widehat{f_2}(x_2)$ }  & \multicolumn{1}{c}{ $\hat{\sigma}_0$}& \multicolumn{1}{c}{ $\rei_{i0}$}  \\
		\cmidrule(l{2pt}r{2pt}){5-6}   \cmidrule(l{2pt}r{2pt}){7-8} \cmidrule(l{2pt}r{2pt}){9-9} \cmidrule(l{2pt}r{2pt}){10-10} 
		Model  & $\sigma_0$  &$\numclusters$ & $\numunits_{i}$  & Bias & Cvg & Bias & Cvg &  Bias & ~RMSEP\\
		\midrule
		GAM & 1 & 100 & 10 & -0.05 & 95 & -0.07 & 95 & -- & --\\
		    &   &     & 20 & -0.04 & 92 & -0.06 & 91 & -- & --\\
		    &   & 200 & 10 & -0.03 & 94 & -0.06 & 95 & -- & --\\
		    &   &     & 20 & -0.02 & 94 & -0.05 & 93 & -- & --\\
		    & 2 & 100 & 10 & -0.05 & 92 & -0.03 & 92 & -- & --\\
		    &   &     & 20 & -0.03 & 88 & -0.03 & 89 & -- & --\\
		    &   & 200 & 10 & -0.04 & 93 & -0.03 & 94 & -- & --\\
		    &   &     & 20 & -0.02 & 89 & -0.02 & 89 & -- & --\\
		\\[-1.8ex]   
		GAMM & 1 & 100 & 10 &  -- & --& -- & -- & -0.05 & 1.62\\
			 &   &     & 20 &  -- & --& -- & -- & -0.02 & 1.76\\
			 &   & 200 & 10 &  -- & --& -- & -- & -0.03 & 1.62\\
			 &   &     & 20 & -- & -- & -- & -- & -0.01 & 1.77\\
			 & 2 & 100 & 10 & -- & -- & -- & -- & -0.08 & 3.46\\
			 &   &     & 20 & -- & -- & -- & -- & -0.05 & 3.64\\
			 &   & 200 & 10 & -- & -- & -- & -- & -0.05 & 3.50\\
			 &   &     & 20 & -- & -- & -- & -- & -0.03 & 3.67\\
		\\[-1.8ex]

		MAM & 1 & 100 & 10 & -0.04 & 96 & -0.01 & 97 & -0.03 & 1.62 \\
			&   &     & 20 & -0.03 & 96 & -0.02 & 96 & -0.01 & 1.76 \\
			&   & 200 & 10 & -0.02 & 96 &  0.00 & 97 & -0.03 & 1.62 \\
			&   &     & 20 & -0.01 & 97 & -0.01 & 95 & -0.01 & 1.77 \\
			& 2 & 100 & 10 & -0.01 & 97 &  0.07 & 96 & -0.04 & 3.46 \\
			&   &     & 20 &  0.00 & 96 &  0.04 & 96 & -0.02 & 3.64 \\
			&   & 200 & 10 &  0.00 & 97 &  0.06 & 96 & -0.04 & 3.49 \\
			&   &     & 20 &  0.01 & 96 &  0.04 & 95 & -0.02 & 3.66 \\
		\hline
	\end{tabular}
\end{table}

\begin{figure}[tbph!]
	\centering
	\subfigure[GAM, $x_{1}$]{\includegraphics[width=0.49\linewidth]{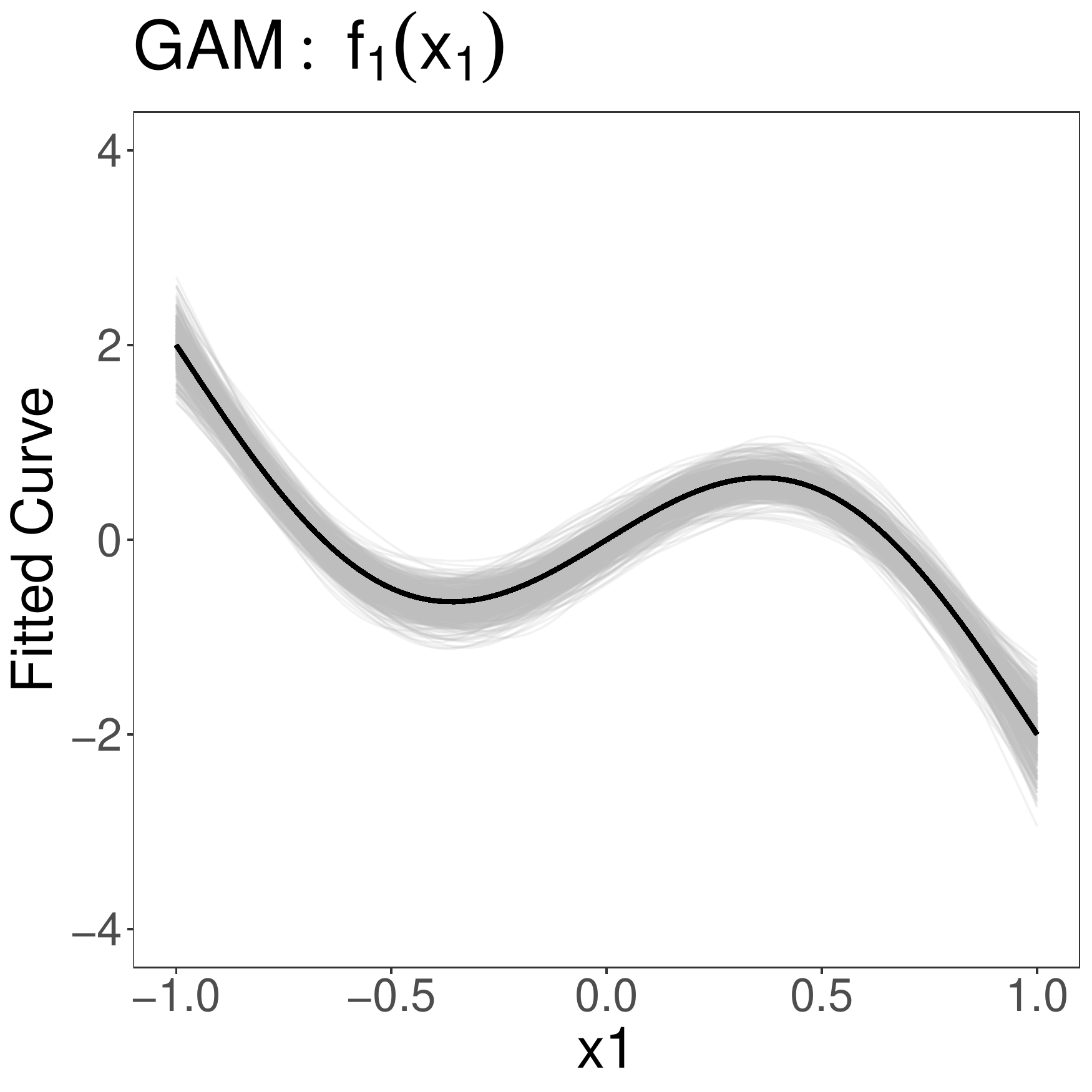}}
	\subfigure[MAM, $x_{1}$]{\includegraphics[width=0.49\linewidth]{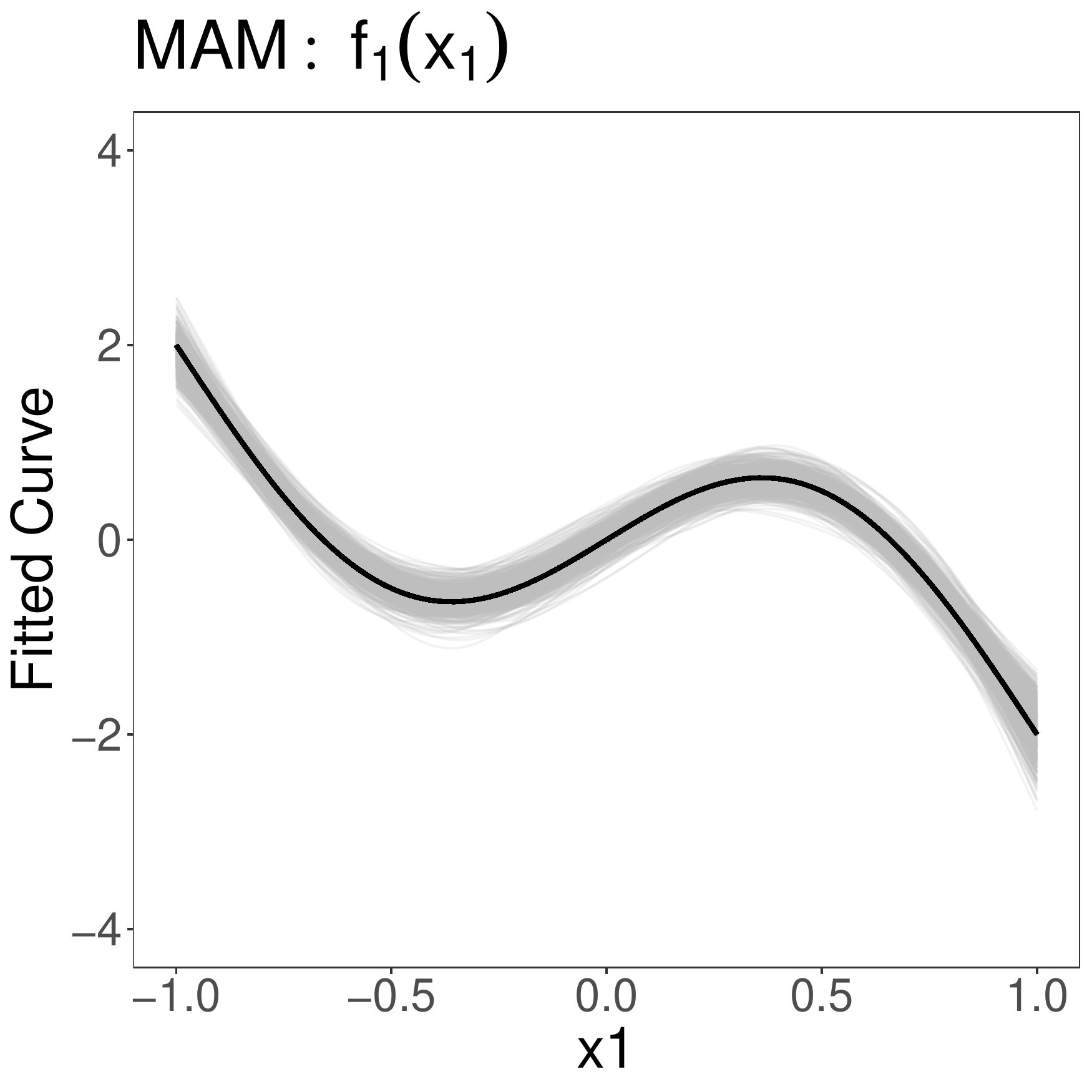}} \\

	\subfigure[GAM, $x_{2}$]{\includegraphics[width=0.49\linewidth]{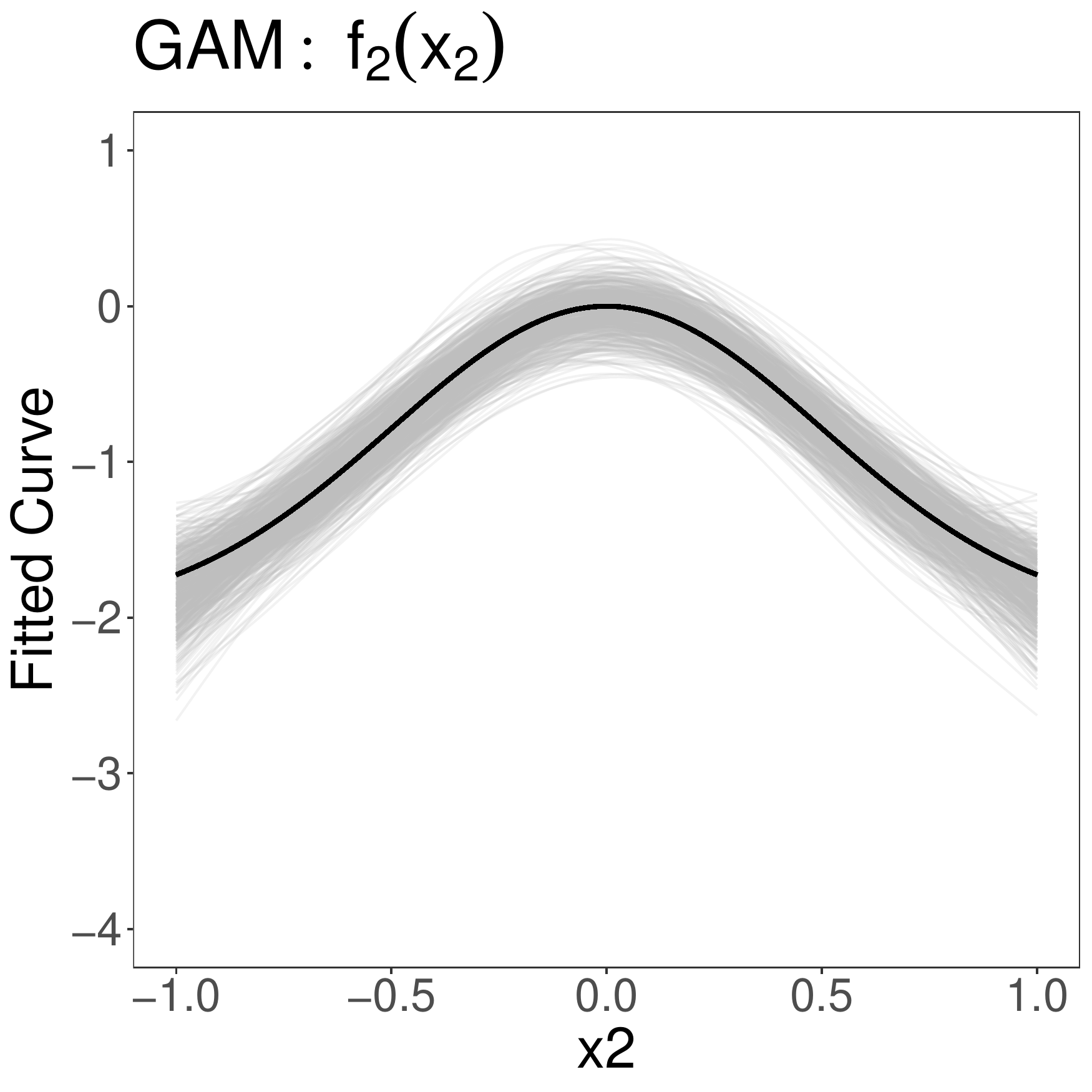}}
	\subfigure[MAM, $x_{2}$]{\includegraphics[width=0.49\linewidth]{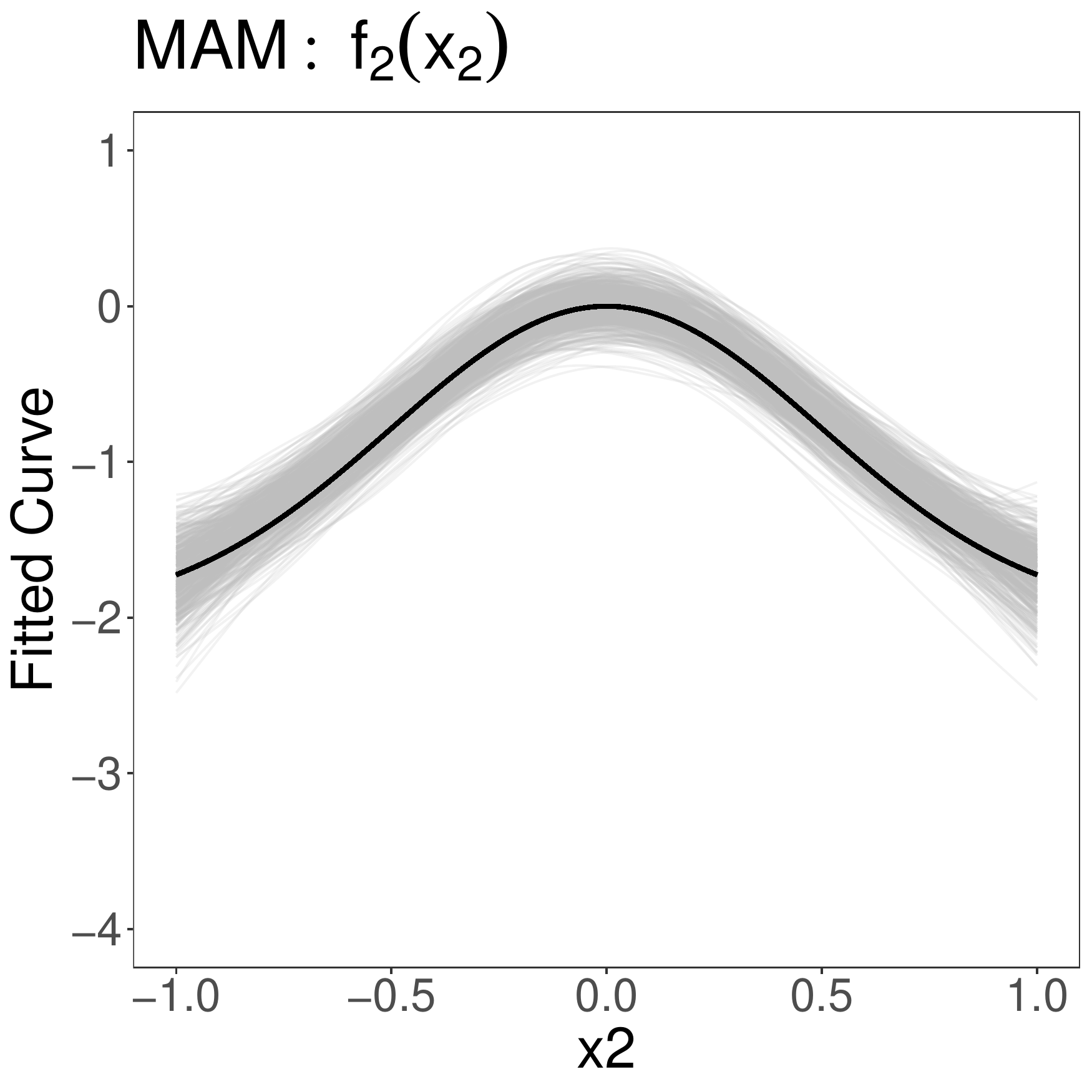}}
	\caption{Simulation results for marginal mean model under random intercepts model with $\numclusters=200$ and $\numunits_{i}=20$. Figure shows distributions of estimates (top row) for  $f_1(\cdot)$ and $f_2(\cdot)$ on a grid of evenly spaced covariate values.}
	\label{fig:sim1int}
\end{figure}

\begin{figure}[tbph!]
	\centering
	\subfigure[GAM, $x_{1}$]{\includegraphics[width=0.49\linewidth]{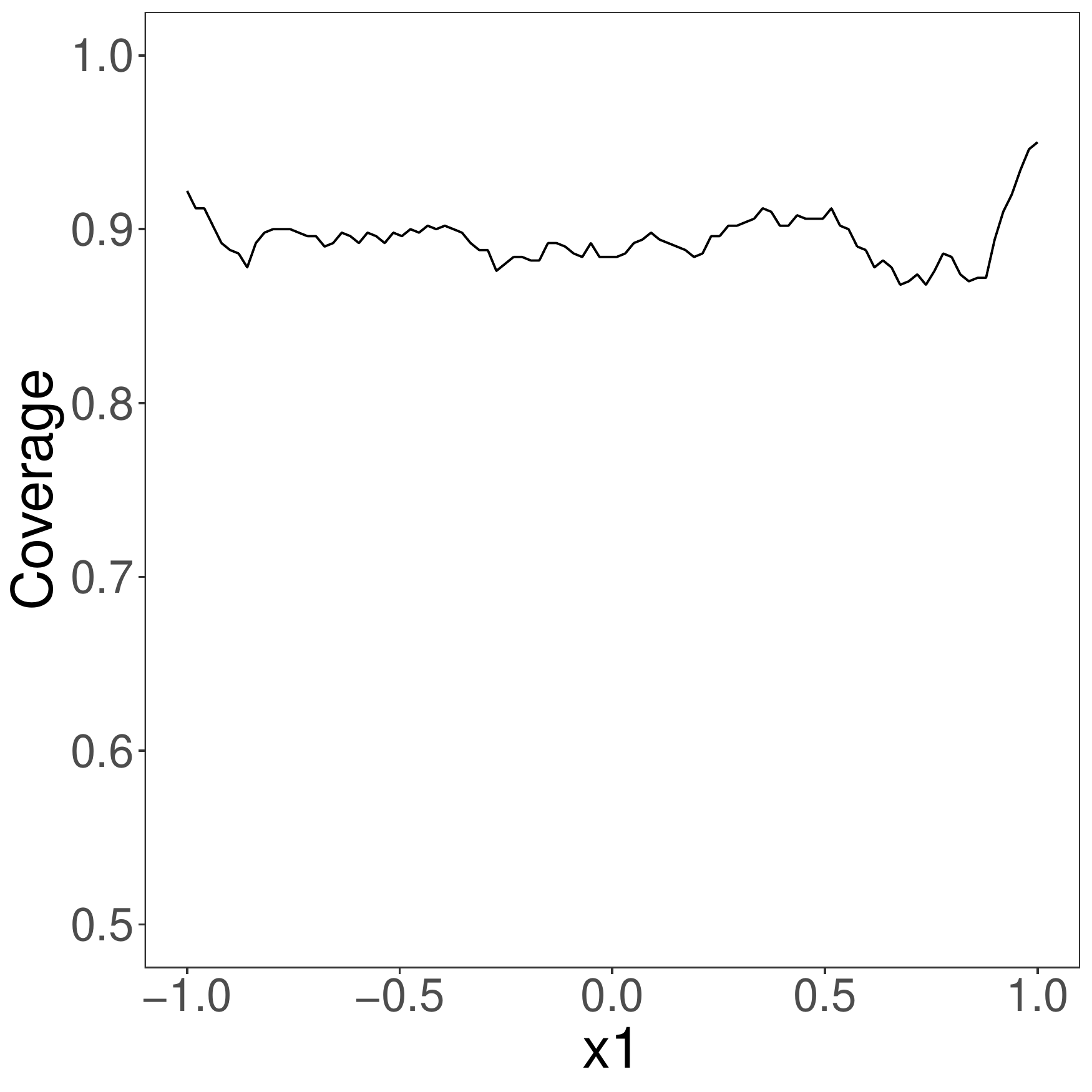}}
	\subfigure[MAM, $x_{1}$]{\includegraphics[width=0.49\linewidth]{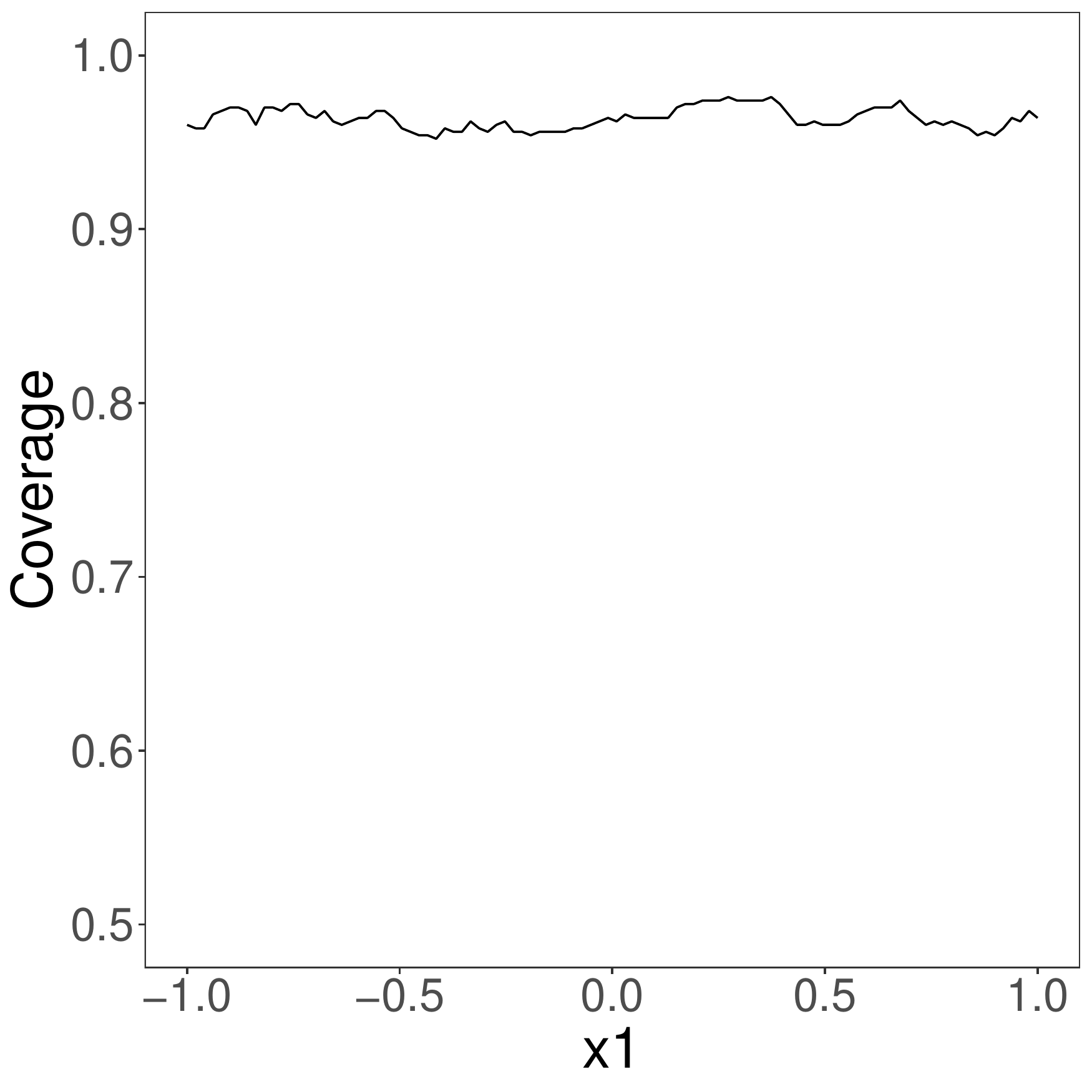}} \\

	\subfigure[GAM, $x_{2}$]{\includegraphics[width=0.49\linewidth]{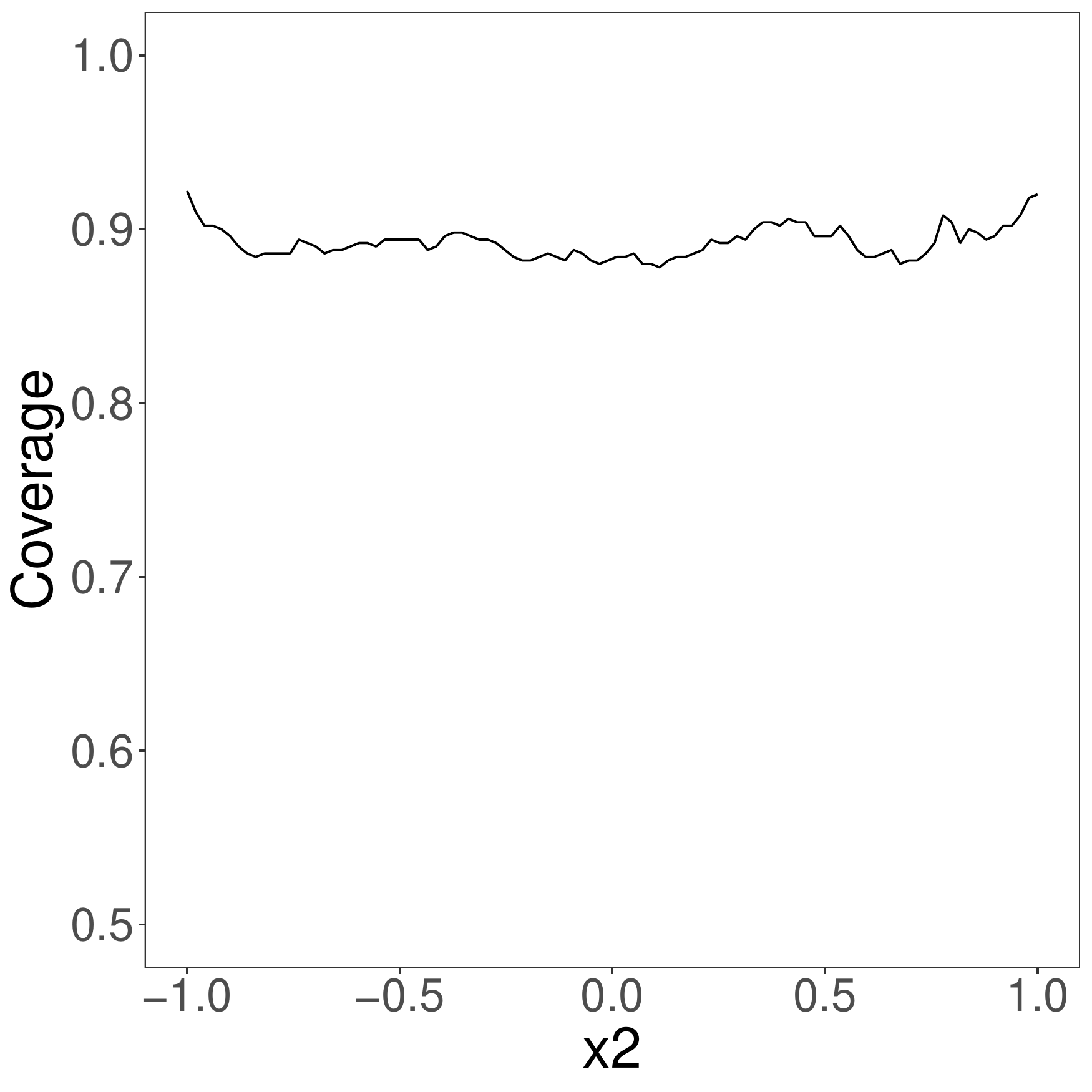}}
	\subfigure[MAM, $x_{2}$]{\includegraphics[width=0.49\linewidth]{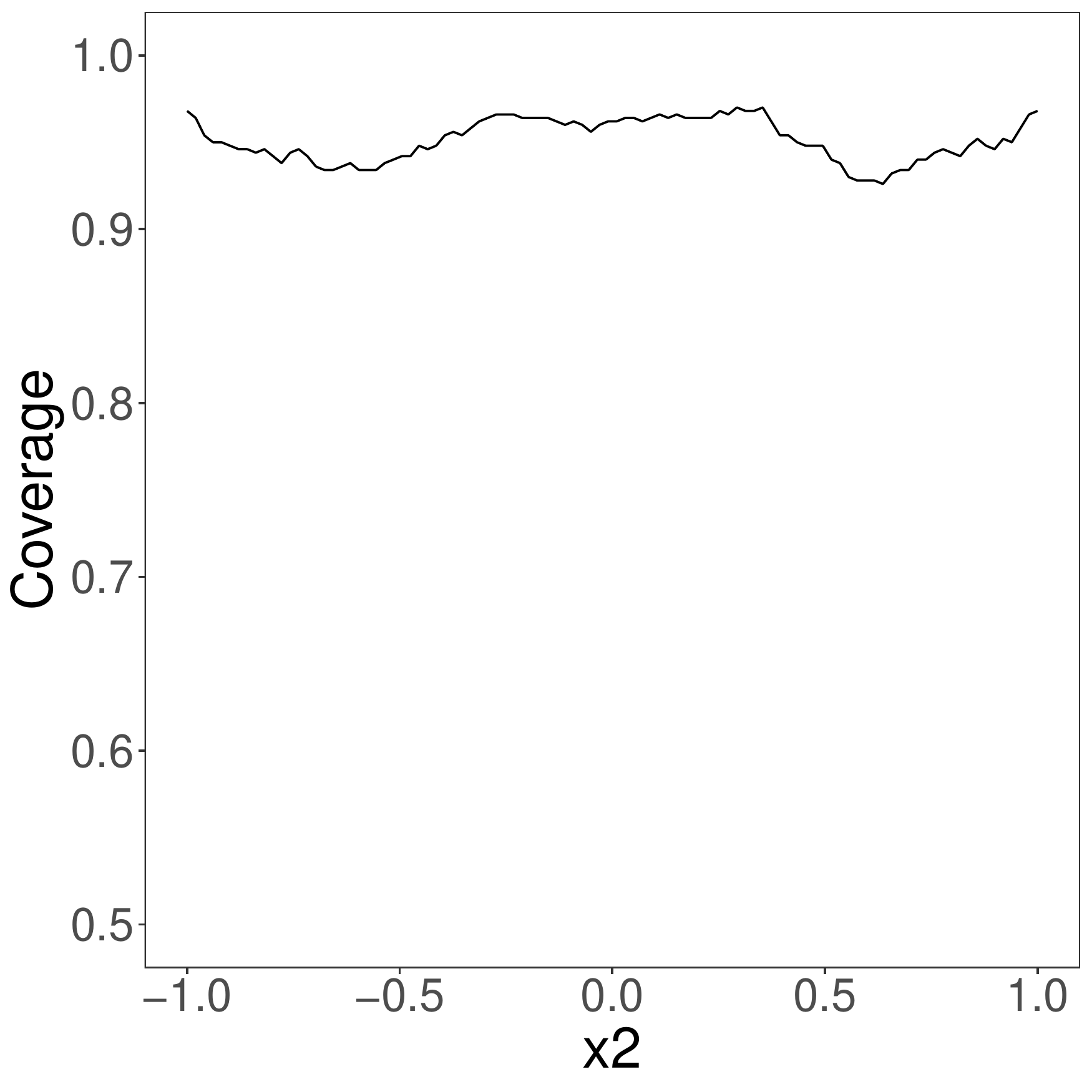}}
	\caption{Simulation results for marginal mean model under random intercepts model with $\numclusters=200$ and $\numunits_{i}=20$. Figure shows confidence interval coverages for  $f_1(\cdot)$ and $f_2(\cdot)$ on a grid of evenly spaced covariate values.}
	\label{fig:sim1intcovr}
\end{figure}

\begin{figure}[tbph!]
	\centering
	\includegraphics[width=1.0\linewidth]{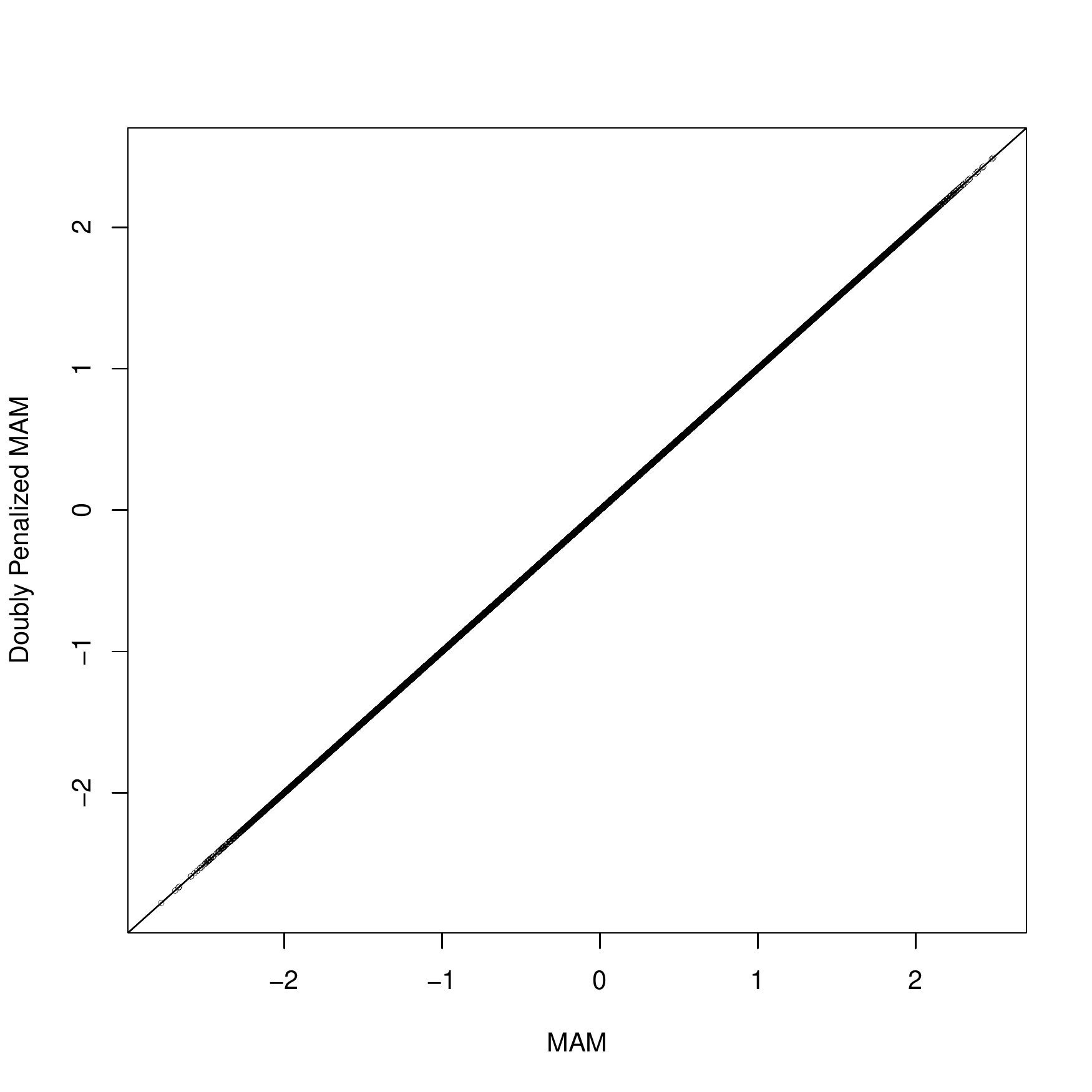}
	\caption{Simulation results for marginal mean model under random intercepts with $\sigma_0=2$, $\numclusters=200$ and $\numunits_{i}=20$. Figure shows fitted values across 500 datasets, comparing the proposed estimation strategy of \S3.2 with an alternative strategy in which marginal estimates are subject to a second penalization:  $\margbasiscoefest_{double}=\text{argmin}\norm{\margmeanlinkest - \margXbasis\margbasiscoef}  +\sum_{l=1}^p \penaltyP_{\penaltyM_{l}}(\margbasiscoef_{l};\penaltymat_{l})$ where $\penaltymat_{l}$ is a $\basisdim_{l}$-dimensional fixed, known penalty matrix with rank $\rank_{l}\leq \basisdim_{l}$ and $\penaltyM_{l}>0$ is a ``smoothing'' parameter to be estimated.}
	\label{fig:doublepenalty}
\end{figure}

\end{document}